\documentclass{aa}

\usepackage{txfonts}
\usepackage{graphicx}
\usepackage{amsmath} 
\usepackage{amssymb}
\usepackage{lscape}
\usepackage{natbib}
\begin{document}
\title{The {\it Herschel} Gould Belt Survey in Chamaeleon~II \thanks{{\it Herschel} is an ESA space observatory with science instruments provided by European-led Principal Investigator consortia and with important participation from NASA.}}

\subtitle{Properties of cold dust in disks around young stellar objects}

\author{L. Spezzi\inst{1} \and  N.L.J. Cox\inst{2} \and T. Prusti\inst{3} \and B. Mer\'in\inst{4} \and \'A. Ribas\inst{4} \and 
C. Alves de Oliveira\inst{5} \and E. Winston\inst{3} \and \'A. K\'osp\'al\inst{3} \and P. Royer\inst{2} \and R. Vavrek\inst{4} \and 
Ph. Andr\'e\inst{6} \and G.L. Pilbratt\inst{3} \and L. Testi\inst{1} \and E. Bressert\inst{7} \and L. Ricci\inst{8}  \and A.  Men'shchikov\inst{7} \and V. K\"onyves\inst{6}}

 \offprints{L. Spezzi, \email{lspezzi@eso.org}}
 
\institute{European Southern Observatory, Karl-Schwarzschild-Strasse 2, 85748 Grating bei M\"unchen, Germany
\and Instituut voor Sterrenkunde, KU\,Leuven, Celestijnenlaan 200D, 3001 Leuven, Belgium
\and Research and Scientifc Support Department, European Space Agency (ESA-ESTEC), PO Box 299, 2200 AG Noordwijk, the Netherlands
\and {\it Herschel} Science Centre, European Space Astronomy Centre (ESA), P.O. Box, 78, 28691 Villanueva de la Ca\~{n}ada, Madrid, Spain 
\and European Space Astronomy Centre (ESA), P.O. Box, 78, 28691 Villanueva de la Ca\~{n}ada, Madrid, Spain 
\and Laboratoire AIM, CEA/DSM-CNRS-Universit Paris Diderot, IRFU/SAp, CEA Saclay, 91191 Gif-sur-Yvette, France
\and CSIRO, Sydney, Australia
\and Division of Physics, Mathematics and Astronomy, California Institute of Technology, MC 249-17, Pasadena, CA, 91125, USA}

\date{Received xxxx ; accepted xxxx}

  \abstract
   {We report on the \emph{Herschel} Gould Belt survey (HGBS) of the Chamaeleon II (Cha~II) star forming region, focusing on the detection 
of Class I to III young stellar objects (YSOs).}
{We aim at characterizing the circumstellar material around these YSOs and understanding which disk parameters are most likely constrained by the new HGBS data, 
expected to be crucial to study the transition from the optically thick disks to the evolved debris-type disks.}
 {We recovered 29 out of the 63 known YSOs in Cha~II with a detection in at least one of the PACS/SPIRE pass-bands: 3 Class I YSOs (i.e.,100\%), 1 Flat source (i.e., 50\%), 21 Class II objects (i.e., 55\%), 
3 Class III objects (i.e, 16\%) and the unclassified far-infrared source IRAS~12522-7640. We explore PACS/SPIRE colors of this sample and present modeling of their spectral energy distributions (SEDs), 
from the optical up to {\it Herschel}'s wavelengths, using the RADMC-2D radiative transfer code.}
{We find that YSO colors are typically confined to the following ranges: $-0.7 \lesssim \log (F_{70} / F_{160}) \lesssim 0.5$, 
$-0.5\lesssim \log (F_{160} / F_{250}) \lesssim 0.6$, $0.05 \lesssim \log (F_{250} / F_{350}) \lesssim 0.25$ and 
$-0.1 \lesssim \log (F_{350} / F_{500}) \lesssim 0.5$.  These color ranges are expected to be only marginally contaminated by extragalactic sources and field stars 
and, hence, provide a useful YSO selection tool when applied altogether. We were able to model the SED of 26 out of the 29 detected YSOs. 
We discuss the degeneracy/limitations of our SED fitting results and adopt the Bayesian method to estimate the probability of different values for the derived disk parameters.
The Cha~II YSOs present typical disk inner radii $\lesssim$0.1~AU, as previously estimated in the literature on the basis of \emph{Spitzer} data. 
Our probability analysis shows that, thanks to the new  \emph{Herschel} data, the lower limits to the disk mass (M$_\mathrm{disk}$) and characteristic radius (R$_\mathrm{C}$) are well constrained, 
while the flaring angle (1+$phi$) is only marginally  constrained. The lower limit to R$_\mathrm{C}$ is typically around 50~AU.
The lower limits to M$_\mathrm{disk}$ are proportional to the stellar masses with a typical 0.3\% ratio, i.e., in the range estimated in the literature for young Class II stars and brown dwarfs 
across a broad range of stellar masses. The estimated flaring angles, although very uncertain, point towards rather flat disks (1+$phi \lesssim$1.2), as found 
for low-mass M-type YSO samples in other star forming regions. Thus, our results support the idea that disk properties show a dependence on stellar properties. 
}
{}


   \maketitle
%
%________________________________________________________________

\section{Introduction}

\begin{figure*}
\centering
\includegraphics[width=15cm]{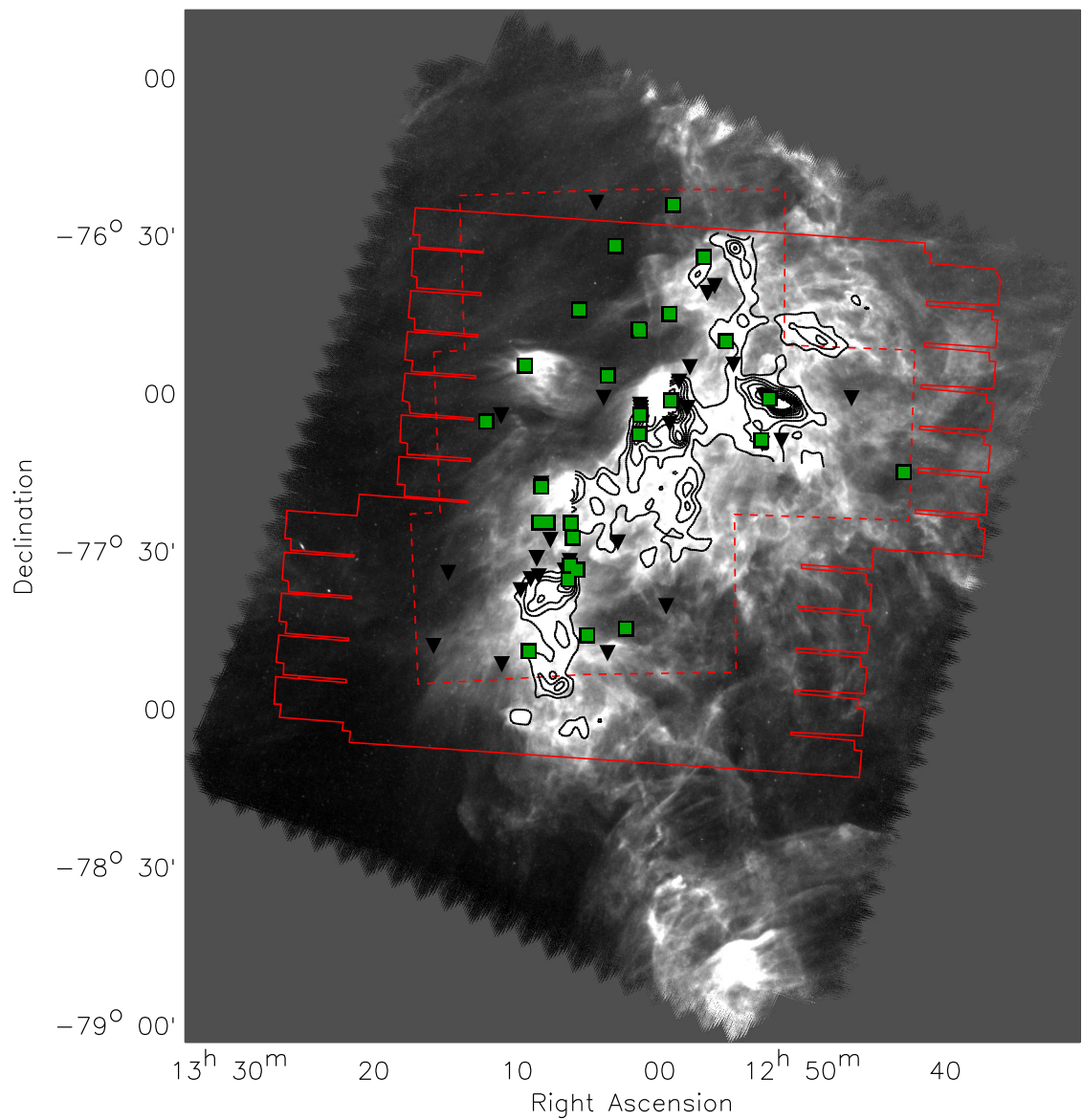}
\caption{SPIRE-250~$\mu$m map of Cha~II. The continuous and dashed lines highlight the areas previously mapped in the infrared with {\it Spitzer} \citep{Alc08} 
and in the optical with ESO-2.2.m/WFI \citep{Spe07}, respectively. The squares are the YSOs detected by the HGBS (Table~\ref{tab_hgb}). The triangles indicate the position of 
previously known YSOs not detected by the HGBS (Table~\ref{tab_UL}). 
The contours from the {\it Spitzer}/IRAC extinction map \citep{Alc08} are also shown.}
\label{spa_distr}
\end{figure*}

Chamaeleon~II (Cha II)  is a low-mass star forming region located in the Chameleon-Musca complex \citep[see][for a recent review]{Luh08}.
Because of its proximity \citep[178$\pm$18~pc;][]{Whi97}, young age \citep[4$\pm$2~Myr;][]{Spe08} 
and compact structure (it extends over less than $\sim$3~deg$^2$ in the sky), 
it has been the target of several multi-wavelength studies over the last $\sim$25 years.
The credit for the observational exploration of the young stellar population of Cha~II goes to many authors and 
a detailed summary of these studies is given in Section~2.1 by \citet{Spe08}. 
Starting from 1977, early objective-prism H$\alpha$ surveys identified the first T~Tauri stars in the region. 
Later on, near to mid-infrared (IR) studies reported the presence of embedded Class I and II sources and 
ROSAT X-ray observations revealed the presence of a number of weak T~Tauri stars.
More recently, deep optical to mid-IR imaging and follow-up spectroscopy observations provided 
a complete census and characterization of the young stellar population in this region down to the brown dwarf regime. 
In 2003, Cha~II was chosen as one of the targets of the {\it Spitzer} Space Telescope Legacy Program 
``From Molecular Cores to Planet-forming Disk`` \citep[c2d;][]{Eva03} and these 
data allowed us to explore, for the first time, the properties of circumstellar disks around the young stellar objects (YSOs) in Cha~II.
The overall results of this survey \citep[][and reference therein]{Alc08} show that the disk fraction in Cha~II (70-80\%) 
is exceptionally high in comparison with other star formation regions of similar age. 
Several of these YSOs show significant mass accretion,  and \citet{Bia12} determined their mass accretion rates.

Altogether, these studies make Cha~II one of the best studied nearby star forming regions and our knowledge of its population 
is comparable to that of other well studied nearby star forming regions, such as Taurus and Orion.
For this reason, Cha~II has been included in the {\it Herschel} Gould Belt survey (HGBS\footnote{http://gouldbelt-herschel.cea.fr}) key project \citep{And10}. 
The ultimate goal of the HGBS is to elucidate the formation mechanisms of prestellar cores out of the diffuse medium, 
crucial for understanding the origin of the distribution of stellar masses, through the observations of cloud filaments, prestellar condensations and Class 0 protostars. 
These questions will be addressed, for the specific case of the Chamaleon complex, in Alves de Oliveira et al. (2013, in preparation). 
Additionally, data from the HGBS can be exploited to study the far-IR and sub-millimeter (sub-mm) emission from more evolved Class I to III YSOs. 
These data are expected to greatly improve the determination of circumstellar disk parameters such as disk mass and degree of flaring \citep{Har12b}, 
which are crucial to understand the multifaceted and still unclear transition from the optically 
thick disks to the evolved debris-type disks, and the associated planet-forming process \citep[see][for a recent review]{Wil11}.
 
In this paper, we focus on the PACS/SPIRE detections and statistics of Class I to III YSOs in Cha~II. Our aim is to characterize the circumstellar 
material around these YSOs and understand which disk parameters 
are reliably constrained using the new HGBS data. This statistical approach is of paramount importance to fully 
exploit {\it Herschel} capabilities in the characterization of disks around T~Tauri stars, because most of the work 
published so far focus on single objects \citep[e.g.,][]{Cie11,Har12a}. The only exception is the paper by \citet{Har12b} which 
analyzes PACS-only data for a sample of  $\sim$50 very low-mass stars and brown dwarfs and, hence, it is biased to lower stellar mass with respect to the sample presented here.
This paper is structured as follows. In Sect.~\ref{sec_obs} we describe the observations and data reduction procedure. 
In Sect.~\ref{sec_yso} we analyze the YSO sample detected by the HGBS in Cha~II.
In Sect.~\ref{sec_cc} we explore the locus of YSOs in SPIRE/PACS color-color diagrams and, in Sect.~\ref{sec_fit}, 
we determine relevant disk parameters for our sample through spectral energy distribution (SED) modeling.
Finally, in Sect.~\ref{concl} we summarize our results and draw our conclusions.

\section{Observations and data reduction \label{sec_obs} }

The Cha~II dark cloud was observed with the {\it Herschel} Space Observatory \citep{Pil10} 
within the frame of the HGBS \citep{And10}. 
The observations (Obs. ID 1342213180 and 1342213181) were conducted on 22-23 January 2011 
in parallel mode using both PACS \citep[Photodetector Array Camera and Spectrometer;][]{Pog10} 
and SPIRE \citep[Spectral and Photometric Imaging REceiver;][]{Gri10}. 
An area of 3.5~deg$^2$, centered at R.A. = 12$^h$58$^m$10$^s$ and Dec. = $-$77$^d$28$\arcmin$28$\arcsec$, 
was covered with PACS at 70 and 160~$\mu$m and with SPIRE at 250, 350 and 500~$\mu$m 
with a scanning speed of 60$\arcsec/s$.  Additionally, PACS only maps at 100~$\mu$m were obtained on the 15 January 2011 (Obs. ID 1342212708, 1342212709) 
with a scanning speed of  20$\arcsec/s$, covering a smaller region of about  2.4~deg$^2$ (Fig.~\ref{spa_distr}). 
The total observing time was $\sim$12~hours, 6~h for the parallel mode and 6~h for the PACS-100~$\mu$m images. 
The observing strategy is described in more detail in \citet{And10}. 

In the case of PACS, the data were reduced using the Scanamorphos map-making software \citep[version 10;][]{Rou12}.
For SPIRE observations we used the \emph{naiveMap} and \emph{destriper} algorithms within HIPE \citep[{\it Herschel} Interactive Programming Environment, version 9;][]{Ott10}. 
The full width at half maximum of the point spread functions (PSFs) indicates that the spatial resolution 
of our maps is  6$\arcsec$, 9$\arcsec$, 12$\arcsec$, 20$\arcsec$, 25$\arcsec$ and 36$\arcsec$ at 70, 100, 160, 250, 350 and 500~$\mu$m, respectively. 
We note that the PSF is slightly elongated in parallel fast scan speed mode \citep{Lut10}.

We visually inspected each map to check the detection of the known YSOs in Cha~II (see Sect.~\ref{sec_yso}), using their coordinates as provided by \citet{Spe08}. 
For the YSOs clearly detected, the photometry (Table~\ref{tab_hgb}) was recovered from the point-source catalog extracted from the PACS/SPIRE maps using 
the multi-wavelength source extraction algorithm \emph{getsources} \citep[][version 1.121206 ] {Men12}; we adopted a searching radius of 15$\arcsec$, 
corresponding to three times the typical FWHM of the sources in our PACS-70$\mu$m map, and visually checked the accuracy of the match for each YSO. 
An aperture correction is applied to the fluxes extracted by \emph{getsources}, taking into account the aperture radii recommended 
for each {\it Herschel} band, i.e., 12$\arcsec$ for 70 and 100~$\mu$m, 22$\arcsec$ for 160 and 250~$\mu$m, 30$\arcsec$ for 350~$\mu$m, and 42$\arcsec$ for 500~$\mu$m 
(see the PACS Point-Source Flux Calibration Technical Note from April 2011, and Sect.~5.7.1.2 of the SPIRE Data Reduction Guide). No color correction is applied. 
When no source was detected, we computed a flux upper limit (Table~\ref{tab_UL}) as in \citet{Rib13}, i.e.,  calculating the RMS of the sky emission over 100 apertures taken around the source, 
using the coordinates provided by \citet{Spe08} and the same aperture radii and correction factors as for detected sources.

\begin{figure}
\centering
\includegraphics[width=7cm]{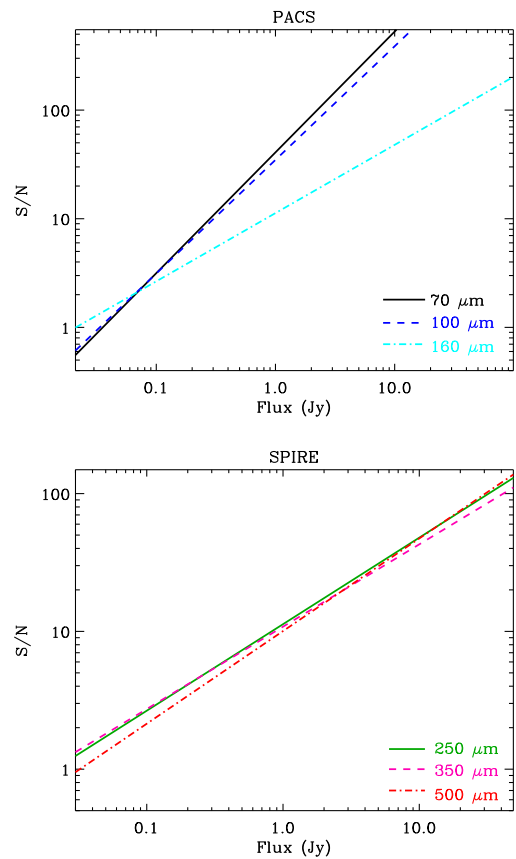}
\caption{Signal to noise ratio as a function of flux for point-like sources in our PACS (upper panel) and SPIRE (lower panel) maps as computed by  \emph{getsources} (version 1.121206) 
in the flux range covered by the Cha~II YSOs.}
\label{SNR}
\end{figure}

Using the multi-wavelength catalog extracted with \emph{getsources}, we also estimated that the 5$\sigma$ detection limit for point-like sources of our maps is $\sim$150~mJy at 70 and 100~$\mu$m 
and $\sim$300~mJy at 160, 250, 350 and 500~$\mu$m (Fig.~\ref{SNR}).

\section{Statistics and fluxes of known YSOs in Cha~II detected by {\it Herschel}  \label{sec_yso}}

The most complete census of YSOs in Cha~II up to date has been presented by \citet{Alc08} and \citet{Spe08}.
The Cha~II young population consists of 63 objects. Their Lada classification \citep{Lad84} is as follows: 
no Class 0 objects, 3 Class I objects, 2 Flat-spectrum sources, 38 Class II objects, 19 Class III objects 
and IRAS~12522-7640, a far-IR source not classified because of the lack of near-IR data. 
We note that classification provided by \citet{Alc08} is based on the SED slope ($\alpha$) of the line joining the flux measurements at 2.2~$\mu$m (K-band)  and MIPS-24~$\mu$m, 
and the Lada class separation as extended by \citet{Gre94}, i.e., $ \alpha \ge 0.3$ for Class I, $-0.3 \le \alpha < 0.3$ for flat-spectrum sources, $-1.6 \le \alpha < -0.3$ 
for Class II sources, and $\alpha < -1.6$ for Class III sources. 

On the basis of our visual inspection of the PACS/SPIRE maps, we found that the HGBS observations detected 29 out of 63 known YSOs in at least one of the six pass-bands: 
3 Class I YSOs (100\% of the previously known Class I sources), 
1 Flat-spectrum source (50\% of the previously known Flat-spectrum sources), 21 Class II objects (55\% of the  previously known Class II population), 
3 Class III object (16\% of the  previously known Class III population) and IRAS~12522-7640. 
Table~\ref{stat_ysos} summarizes the statistics of HGBS YSO detections in Cha~II. 
In Table~\ref{tab_hgb} we report the PACS and SPIRE fluxes for these 29 objects, together with their spectra type, Lada class, HGBS coordinates and 
the distance between the HGBS coordinates and the coordinates reported by \citet{Spe08} and \citet{Alc08}; we note that this distance varies from a few to a few-tenth arcsec, 
in agreement with the spatial resolution of the HGBS maps (Sect.~\ref{sec_obs}).

For the sake of completeness, we report in Table~\ref{tab_UL} the flux upper limits in each band for the 34 undetected YSOs. 
We cannot add new information on these sources and, hence, do not treat them further in this paper. 
However, we notice that the estimated upper limits are consistent with the Lada classification and the SEDs presented for these objects by \citet{Alc08}.

\begin{figure}
\centering
\includegraphics[width=9cm]{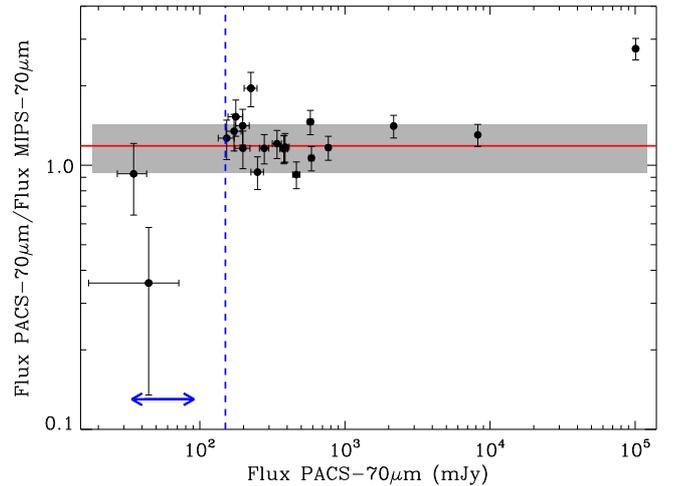}
\caption{Comparison between the flux at 70~$\mu$m as measured by PACS and the flux measured by MIPS at the same wavelength 
for the 23 YSOs in Cha~II detected by the HGBS. The solid line and shaded area indicate the median value and the RMS of the flux ratio, respectively. 
The arrow-head line displays the MIPS-70~$\mu$m flux range of Cha~II YSOs not detected by PACS. 
The dashed line indicates the PACS-70~$\mu$m 5$\sigma$ detection limit. The clear outlier is Sz~49, the faintest YSO detected by PACS.}
\label{70micr_comparison}
\end{figure}

As an additional check of our data reduction and flux extraction procedures, we also compared the 70~$\mu$m-flux 
of the detected YSOs as measured by PACS with the flux measured at the same wavelength with {\it Spitzer}/MIPS \citep{Alc08}. 
In Fig.~\ref{70micr_comparison} we show the ratios of the PACS and MIPS fluxes at 70~$\mu$m as a function of the PACS fluxes. 
We find a typical value of 1.18 for this ratio and a RMS of 0.24, meaning that there is no significant shift between the two datasets. 
Considering the uncertainties on PACS and MIPS flux measurements, the only clear outlier is IRAS~12496-7650, for which the PACS-70~$\mu$m flux 
is almost three times larger than the MIPS-70~$\mu$m one; we visually inspected this source in the PACS/MIPS  mosaics and found out that its MIPS-70~$\mu$m flux \citep[36.5~Jy;][]{Alc08}
is slightly above the saturation limit of the MIPS mosaic, where saturation for point-like sources in low backgrounds occurs around 23~Jy \citep{Eva07}.

Moreover, Fig.~\ref{70micr_comparison} shows that we basically detect with PACS all YSOs with flux density greater than $\sim$100~mJy in MIPS-70~$\mu$m.
This value is slightly lower than the 5$\sigma$ detection limit of our PACS-70~$\mu$m photometry (Sect.~\ref{sec_obs}).

\begin{table}
\centering
\caption{ of Class I to III YSOs detected by the HGBS in Cha~II. We report between brackets the percentage of detections with respect to the total number of known YSOs in the given class.}           
\label{stat_ysos}                           
\begin{tabular}{ll}      
\hline\hline               
 Lada  Class       & N. of detections \\
\hline  
  I                & 3 (100\%) \\
 Flat           & 1 (50\%) \\
 II               & 21 (55\%) \\   
 III              & 3 (16\%) \\
 Unknown & 1 (100\%)\\
\hline                                   
\end{tabular}
\\
\end{table}

\begin{table*}
\caption{{\it Herschel} PACS (70 to 160 $\mu$m) and SPIRE (250 to 500 $\mu$m) fluxes for known YSOs in Cha~II detected by the HGBS.}           
\label{tab_hgb}                           
\tiny\addtolength{\tabcolsep}{-4pt}
\begin{tabular}{llllllllllll}      
\hline\hline               
Main                 &  Spectral     & Lada           & RA(J2000)$^\star$ & DEC(J2000)$^\star$ & Dist.$^\bigtriangleup$       & 70~$\mu$m$^\spadesuit$ &   100~$\mu$m$^\spadesuit$ & 160~$\mu$m$^\spadesuit$ &  250~$\mu$m$^\spadesuit$ &  350$\mu$m$^\spadesuit$  & 500~$\mu$m$^\spadesuit$ \\ 
Designation   &     Type$^\dag$      & Class$^\ddag$  &   (hh$:$mm$:$ss)  &     (dd$:$mm$:$ss) &  ($\arcsec$)  &   (Jy)                         &  (Jy)                   &     (Jy)               &     (Jy)                        &    (Jy)                           &    (Jy)      \\
\hline                        
       IRAS~12416-7703 &    M2.5   	     &    II   &      12:45:07.85  &	-77:20:12.29 &    4.8	&     0.18$\pm$0.02   &    OOF$^\amalg$ 	   &	$<$9.00 	&	  $<$0.02	  &	   $<$0.13	   &	$<$0.15  \\ 
       IRAS~12496-7650 &    F0      	     &    II   &      12:53:17.12  &	-77:07:10.86 &    0.3	&   100.80$\pm$0.19   &   79.21$\pm$0.20   &   46.16$\pm$0.59	&	 24.84$\pm$0.38   &	  15.14$\pm$0.62   &   11.82$\pm$0.32  \\ 
       IRAS~12500-7658 &    K5$^\clubsuit$   &     I   &      12:53:42.83  &	-77:15:11.78 &    0.2	&     2.17$\pm$0.03   &    2.44$\pm$0.02   &	2.07$\pm$0.05	&	  1.50$\pm$0.06   &	   0.98$\pm$0.07   &	0.49$\pm$0.13  \\ 
       IRAS~12522-7640 &   --	      	     &    --   &      12:55:50.06  &	-76:56:10.03 &   10.8	&     1.00$\pm$0.04   &    0.84$\pm$0.05   &	0.81$\pm$0.10	&	  0.98$\pm$0.19   &	   0.93$\pm$0.20   &	0.48$\pm$0.08  \\ 
       IRAS~12535-7623 &    M0      	     &    II   &      12:57:11.96  &	-76:40:10.02 &    1.6	&     0.20$\pm$0.02   &    0.13$\pm$0.01   &	0.13$\pm$0.03	&	  0.07$\pm$0.06   &	   $<$0.04	   &	$<$0.02  \\ 
 WFI~J12585611-7630105 &    M5      	     &   III   &      12:58:55.62  &	-76:30:10.77 &    1.7	&     0.15$\pm$0.02   &    0.14$\pm$0.01   &	0.05$\pm$0.03	&	  0.06$\pm$0.03   &	   $<$0.03	   &	$<$0.03  \\ 
	  ISO-CHAII~28 &  K4.3      	     &     I   &      12:59:06.68  &	-77:07:40.41 &    0.5	&     8.23$\pm$0.04   &    6.69$\pm$0.03   &	5.31$\pm$0.10	&	 11.71$\pm$0.30   &	  14.10$\pm$0.58   &   33.70$\pm$0.49  \\ 
		  C~41 &  M5.5      	     &     F   &      12:59:11.39  &	-76:51:02.91 &    5.3	&     0.17$\pm$0.02   &    0.11$\pm$0.02   &	0.44$\pm$0.06	&	  0.13$\pm$0.07   &	   0.07$\pm$0.06   &	$<$0.03  \\ 
		 Sz~49 &  M0.5               &    II   &      13:00:53.51  &	 -76:54:14.32 &   1.1	&      0.04$\pm$0.03   &    0.18$\pm$0.01   &	 0.15$\pm$0.03   &	   0.14$\pm$0.02   &	    0.11$\pm$0.03   &	 0.07$\pm$0.03  \\ 
		 Sz~50 &    M3               &    II   &      13:00:55.46  &	 -77:10:23.25 &   1.4	&      0.59$\pm$0.02   &    0.55$\pm$0.02   &	 0.31$\pm$0.07   &	   $<$0.15	   &	    $<$0.03	    &	 $<$0.01  \\ 
      RX~J1301.0-7654a &    K5               &   III   &      13:00:56.56  &    -76:54:04.45 &	  2.9	&    0.02$\pm$0.01   &    0.01$\pm$0.00   &    $<$0.14         &	 $<$0.03	 &	  $<$0.03	  &    $<$0.02  \\ 
      IRASF~12571-7657 &      K3             &    II   &      13:00:59.60  &	-77:14:02.95 &    1.3	&     0.38$\pm$0.02   &    0.31$\pm$0.02   &	0.24$\pm$0.05	&	  $<$0.10	  &	   $<$0.06	   &	$<$0.03  \\ 
		 Sz~51 &    K8.5             &    II   &      13:01:58.33  &	-77:51:22.76 &    2.2	&     0.15$\pm$0.02   &    0.10$\pm$0.01   &	0.29$\pm$0.06	&	  0.04$\pm$0.02   &	   $<$0.02	   &	$<$0.02  \\ 
		CM~Cha &      K7             &    II   &      13:02:13.71  &	-76:37:57.11 &    0.9	&     0.58$\pm$0.03   &    0.46$\pm$0.03   &	0.43$\pm$0.02	&	  0.43$\pm$0.02   &	   0.38$\pm$0.01   &	0.26$\pm$0.01  \\ 
       IRAS~12589-7646 &    M4               &    III  &      13:02:50.60  &    -77:02:47.46 &	  9.7	&    $<$0.03	     &    1.73$\pm$0.00   &    0.08$\pm$0.03   &	 $<$0.02	 &	  $<$0.02	  &    $<$0.02  \\ 
		 Hn~22 &    M2               &    II   &      13:04:23.57  &    -76:50:04.27 &	  3.1	&    0.39$\pm$0.03   &    0.39$\pm$0.02   &    $<$0.48         &	 $<$0.37	 &	  $<$0.28	  &    $<$0.19  \\ 
		 Hn~23 &    K5               &    II   &      13:04:23.94  &    -76:50:03.85 &	  2.5	&    0.38$\pm$0.02   &    0.35$\pm$0.03   &    $<$0.48         &	 $<$0.37	 &	  $<$0.28	  &    $<$0.19  \\ 
		 Sz~52 &   M2.5              &    II   &      13:04:24.93  &    -77:52:31.46 &	  1.4	&    $<$0.18	     &    0.04$\pm$0.02   &    $<$0.90         &	 $<$0.09	 &	  $<$0.06	  &    $<$0.10  \\ 
		 Hn~24 &    M0   	     &    II   &      13:04:55.94  &	-77:39:50.41 &    1.4	&     0.28$\pm$0.02   &    0.25$\pm$0.01   &	0.25$\pm$0.07	&	  $<$0.15	  &	   $<$0.09	   &	$<$0.01  \\ 
		 Hn~25 &  M2.5   	     &    II   &      13:05:08.47  &	-77:33:42.58 &    0.1	&     0.22$\pm$0.02   &    0.23$\pm$0.01   &	0.19$\pm$0.03	&	  0.13$\pm$0.06   &	   $<$0.02	   &	$<$0.01  \\ 
		 Sz~53 &    M1   	     &    II   &      13:05:12.68  &	-77:30:52.68 &    0.1	&     0.20$\pm$0.02   &    0.15$\pm$0.02   &	0.03$\pm$0.03	&	  0.07$\pm$0.04   &	   $<$0.04	   &	$<$0.01  \\ 
		 Sz~54 &    K5   	     &    II   &      13:05:20.99  &	-77:39:00.91 &    0.8	&     0.34$\pm$0.02   &    0.28$\pm$0.02   &	0.24$\pm$0.07	&	  $<$0.10	  &	   $<$0.03	   &	$<$0.01  \\ 
STc2d~J130529.0-774140 &    --		     &    II   &      13:05:27.08  &	-77:41:21.51 &   14.6	&	  $<$0.43	  &    0.04$\pm$0.02   &    $<$0.31 &	      $<$0.16	      &        $<$0.24         &    $<$0.30  \\ 
		 Sz~56 &    M4               &    II   &      13:06:38.69  &    -77:30:34.47 &	  0.9	&    0.04$\pm$0.01   &    0.02$\pm$0.00   &    $<$0.02 &	 $<$0.06	 &	  $<$0.06	  &    $<$0.05  \\ 
		 Sz~58 &  K5   		     &    II   &     13:06:57.29  &    -77:23:41.16 &	  0.4	&    0.77$\pm$0.02   &    0.91$\pm$0.02   &    1.01$\pm$0.05   &	 0.83$\pm$0.09   &	  0.67$\pm$0.11   &    0.43$\pm$0.11  \\ 
		 Sz~59 &  K7   		     &    II   &     13:07:09.10  &    -77:30:30.38 &	  0.4	&    0.25$\pm$0.02   &    0.32$\pm$0.01   &    0.16$\pm$0.03   &	 0.11$\pm$0.03   &	  $<$0.04	  &    $<$0.04  \\ 
       IRAS~13036-7644 &  --   		     &I$^\diamondsuit$&13:07:37.02&    -77:00:26.17 &	  6.1	&    8.11$\pm$0.06   &    OOF$^\amalg$  	  &    9.83$\pm$0.64   &	32.77$\pm$0.49   &	 30.93$\pm$0.43   &   19.72$\pm$0.24  \\ 
		 Sz~61 &  K5   		     &    II   &     13:08:06.41  &    -77:55:05.37 &	  0.4	&    0.46$\pm$0.03   &    0.45$\pm$0.02   &    0.49$\pm$0.03   &	 0.45$\pm$0.03   &	  0.46$\pm$0.03   &    0.41$\pm$0.08  \\ 
		 Sz~63 &  M3   		     &    II   &     13:10:04.32  &    -77:10:43.99 &	  0.9	&    0.21$\pm$0.02   &    OOF$^\amalg$  	  &    0.36$\pm$0.02   &	 0.25$\pm$0.01   &	  0.19$\pm$0.01   &    0.13$\pm$0.01  \\ 
 \hline                                   
\end{tabular}
\\
$^\dag$ From \citet{Spe08}.\\
$^\ddag$ From \citet{Alc08}.\\
$^\star$ Coordinates are from the PACS-70~$\mu$m map. For IRAS~12589-7646, Sz~52 and STc2d~J130529.0-774140 coordinates from the PACS-100~$\mu$m map are reported.\\
$^\bigtriangleup$ Angular distance between the HGBS coordinates and the coordinates reported by \citet{Spe08} and \citet{Alc08}.\\
$^\spadesuit$ The reported uncertainties are flux extraction errors from  \emph{getsources}. The absolute calibration errors for PACS and SPIRE are 5\% and 7\%, respectively (see PACS and SPIRE observer manuals). \\
$^\clubsuit$ Estimated by J.M. Alcal\'a (private communication).\\
$^\diamondsuit$ From \citet{Leh05}.\\
$^\amalg$ Out Of Field (see Sect.~\ref{sec_obs}).\\
\end{table*}

\begin{table*}
\caption{Flux upper limits for known YSOs in Cha~II not detected by the HGBS.}           
\label{tab_UL} 
\tiny                          
\begin{tabular}{lllllcccccc}      
\hline\hline               
Main                &  Spectral                  & Lada                        & RAJ2000$^\star$            & DECJ2000$^\star$ &   70~$\mu$m &   100~$\mu$m & 160~$\mu$m &  250~$\mu$m &  350$\mu$m  & 500~$\mu$m \\ 
 Designation   &     Type$^\dag$       &    Class$^\dag$    &           (hh$:$mm$:$ss)      &       (dd$:$mm$:$ss)            &         (Jy)                         &  (Jy)                   &     (Jy)               &     (Jy)                        &    (Jy)                           &    (Jy)      \\
\hline
           IRAS~12448-7650 & M0.5   &  III    & 12:48:25.70 &    -77:06:36.72 &	    $<$0.04	&   $<$0.03   &   $<$0.03    &   $<$0.04   &   $<$0.04   &    $<$0.03  \\ 
          IRASF~12488-7658 & M5.5   &   III   & 12:52:30.49 &    -77:15:12.92 &	    $<$0.01	&   $<$0.01   &   $<$0.01    &   $<$0.05   &   $<$0.05   &    $<$0.04  \\ 
     WFI~J12533662-7706393 & M6     &  III    & 12:53:36.62 &    -77:06:39.31 &	    $<$0.01	&   $<$0.07   &   $<$0.07    &   $<$0.19   &   $<$0.23   &    $<$0.34  \\ 
                      C~17 & M1.5   &  III    & 12:53:38.84 &    -77:15:53.21 &	    $<$0.01	&   $<$0.01   &   $<$0.07    &   $<$0.07   &   $<$0.10   &    $<$0.10  \\ 
                      C~33 & M1     &  III    & 12:55:25.72 &    -77:00:46.62 &	    $<$0.01	&   $<$0.01   &   $<$0.08    &   $<$0.10   &   $<$0.12   &    $<$0.14  \\ 
                    Sz~46N & M1     &   II    & 12:56:33.59 &    -76:45:45.18 &	    $<$0.08	&   $<$0.01   &   $<$0.18    &   $<$0.05   &   $<$0.04   &    $<$0.03  \\ 
                     Sz~47 & --     &  III    & 12:56:58.63 &    -76:47:06.72 &	    $<$0.00	&   $<$0.00   &   $<$0.00    &   $<$0.06   &   $<$0.08   &    $<$0.08  \\ 
   SSTc2d~J125758.7-770120 & M9     &   II    & 12:57:58.70 &    -77:01:19.50 &	    $<$0.02	&   $<$0.02   &   $<$0.05    &   $<$0.04   &   $<$0.05   &    $<$0.04  \\ 
              ISO-CHAII~13 & M7     &   II    & 12:58:06.67 &    -77:09:09.22 &	    $<$0.07	&   $<$0.10   &   $<$0.10    &   $<$0.25   &   $<$0.28   &    $<$0.20  \\ 
     WFI~J12583675-7704065 & M9     &  III    & 12:58:36.75 &    -77:04:06.53 &	    $<$0.00	&   $<$0.02   &   $<$0.01    &   $<$0.05   &   $<$0.07   &    $<$0.10  \\ 
              ISO-CHAII~29 & M0     &  III    & 12:59:10.19 &    -77:12:13.72 &	    $<$0.05	&   $<$0.09   &   $<$0.59$^\clubsuit$    &   $<$1.34$^\clubsuit$   &   $<$2.24$^\clubsuit$   &    $<$2.65$^\clubsuit$  \\ 
           IRAS~12556-7731 & M5     &  III    & 12:59:26.50 &    -77:47:08.70 &	    $<$0.90	&   $<$0.36   &   $<$0.05    &   $<$0.16   &   $<$0.15   &    $<$0.09  \\ 
     WFI~J13005297-7709478 & M9     &  III    & 13:00:52.97 &    -77:09:47.77 &	    $<$0.10	&   $<$0.03   &   $<$0.20    &   $<$0.20   &   $<$0.20   &    $<$0.30  \\ 
          Sz~48NE$^\ddag$  & M0.5   &	II    & 13:00:53.15 &	 -77:09:09.18 &     $<$0.02	&   $<$0.03   &   $<$0.07    &   $<$0.23   &   $<$0.31   &    $<$0.28  \\ 
          Sz~48SW$^\ddag$  & M1     &	II    & 13:00:53.56 &	 -77:09:08.28 &     $<$0.02	&   $<$0.03   &   $<$0.07    &   $<$0.23   &   $<$0.31   &    $<$0.28  \\ 
     WFI~J13005531-7708295 & M2.5   &  III    & 13:00:55.31 &    -77:08:29.54 &	    $<$0.00	&   $<$0.00   &   $<$0.00    &   $<$0.19   &   $<$0.27   &    $<$0.57  \\ 
                      C~50 & M5     &   II    & 13:02:22.82 &    -77:34:49.51 &	    $<$0.00	&   $<$0.00   &   $<$0.03    &   $<$0.09   &   $<$0.10   &    $<$0.10  \\ 
           RX~J1303.1-7706 & M0     & III     & 13:03:04.46 &    -77:07:02.75 &	    $<$0.02	&   $<$0.01   &   $<$0.01    &   $<$0.02   &   $<$0.02   &    $<$0.03  \\ 
                      C~51 & M4.5   &  III    & 13:03:09.04 &    -77:55:59.52 &	    $<$0.01	&   $<$0.01   &   $<$0.01    &   $<$0.05   &   $<$0.06   &    $<$0.06  \\ 
     WFI~J13031615-7629381 & M7     &  III    & 13:03:16.15 &    -76:29:38.15 &	    $<$0.00	&   OOF$^\amalg$       &   $<$0.01    &   $<$0.01   &   $<$0.04   &    $<$0.01  \\ 
   SSTc2d~J130521.7-773810 & C      &    F    & 13:05:21.66 &    -77:38:10.14 &	    $<$0.02	&   $<$0.01   &   $<$0.01    &   $<$0.02   &   $<$0.02   &    $<$0.20  \\ 
   SSTc2d~J130540.8-773958 & L1     &   II    & 13:05:40.80 &    -77:39:58.20 &	    $<$0.00	&   $<$0.01   &   $<$0.01    &   $<$0.16   &   $<$0.22   &    $<$0.29  \\ 
                     Sz~55 & M2     &   II    & 13:06:30.49 &    -77:34:00.12 &	    $<$0.02	&   $<$0.05   &   $<$0.01    &   $<$0.04   &   $<$0.03   &    $<$0.08  \\ 
                     Sz~57 & M5     &   II    & 13:06:56.56 &    -77:23:09.46 &	    $<$0.05	&   $<$0.90   &   $<$0.14    &   $<$0.03   &   $<$0.04   &    $<$0.04  \\ 
                      C~62 & M4.5   &   II    & 13:07:18.04 &    -77:40:53.00 &	    $<$0.04	&   $<$0.02   &   $<$0.01    &   $<$0.12   &   $<$0.20   &    $<$0.19  \\ 
            Sz~60W$^\ddag$ & M1     &  III    & 13:07:22.30 &	 -77:37:22.62 &     $<$0.04	&   $<$0.06   &   $<$0.06    &   $<$0.03   &   $<$0.02   &    $<$0.02  \\ 
            Sz~60E$^\ddag$ & M4     &	II    & 13:07:23.33 &	 -77:37:23.20 &     $<$0.04	&   $<$0.06   &   $<$0.06    &   $<$0.03   &   $<$0.02   &    $<$0.02  \\ 
                     Hn~26 & M2     &   II    & 13:07:48.50 &    -77:41:21.73 &	    $<$0.02	&   $<$0.05   &   $<$0.13    &   $<$0.09   &   $<$0.12   &    $<$0.13  \\ 
                      C~66 & M4.5   &   II    & 13:08:27.19 &    -77:43:23.41 &	    $<$0.03	&   $<$0.03   &   $<$0.10    &   $<$0.13   &   $<$0.48   &    $<$0.66  \\ 
IRASF~13052-7653NW$^\ddag$ & M0.5   &	II    & 13:09:09.81 &	 -77:09:43.52 &     $<$0.04	&   OOF$^\amalg$       &   $<$0.07    &   $<$0.02   &	$<$0.01   &    $<$0.02  \\ 
 IRASF~13052-7653N$^\ddag$ & M1.5   &	II    & 13:09:10.98 &	 -77:09:44.14 &     $<$0.04	&   OOF$^\amalg$       &   $<$0.07    &   $<$0.02   &	$<$0.01   &    $<$0.02  \\ 
                     Sz~62 & M2.5   &   II    & 13:09:50.44 &    -77:57:23.94 &	    $<$0.02	&   $<$0.06   &   $<$0.08    &   $<$0.06   &   $<$0.05   &    $<$0.03  \\ 
    2MASS~13125238-7739182 & M4.5   &  III    & 13:12:52.37 &    -77:39:18.58 &	    $<$0.03	&   OOF$^\amalg$       &   $<$0.01    &   $<$0.01   &   $<$0.01   &    $<$0.01  \\ 
                     Sz~64 & M5     &   II    & 13:14:03.83 &    -77:53:07.48 &	    $<$0.03	&   $<$0.03   &   $<$0.10    &   $<$0.03   &   $<$0.03   &    $<$0.03  \\ 
\hline                                   
\end{tabular}
\\
$^\dag$ Spectral types and Lada classes are from \citet{Spe08} and \citet{Alc08}, respectively.\\
$^\star$ Coordinates are from the optical survey by \citet{Spe08}.\\
$^\clubsuit$ The object is located in a region with strong background emission. \\
$^\ddag$ Binary system \citep{Spe08}.\\
$^\amalg$ Out Of Field (see Sect.~\ref{sec_obs}).\\

\end{table*}

\section{Exploring the YSO locus in PACS/SPIRE color-color diagrams \label{sec_cc}}

Infrared color-color (CC) diagrams are traditional diagnostic tools for the investigation of circumstellar matter around YSOs \citep[][and references therein]{Har05,Lad06}. 
However, the identification/classification of YSOs on the basis of mid to far-IR colors is not trivial because of the high level of contamination \citep[Sect.~3.1 by][]{Eva09,Har07,Oli09}.
Indeed,  YSO colors in this wavelength regime are very similar to those of many background galaxies and, to a smaller extent, 
may also be mimicked by highly reddened stellar photospheres of older field stars. 
\citet{Har07} provided a YSO selection tool on the basis of {\it Spitzer} IRAC/MIPS colors. By using {\it Spitzer} observations of the Serpens star forming region 
and the SWIRE catalog of extragalactic sources  \citep{Lon03}, these authors defined the boundaries of the YSO locus in several IRAC/MIPS color-magnitude and CC diagrams. Their criteria have proven to provide an optimal separation between young stars and galaxies, with the fraction of remaining contaminants estimated to be around 30\% \cite[e.g.,][]{Spe08,Oli09,Cie10}, and have been applied to select YSO candidates in all star forming regions observed within the frame of the {\it Spitzer} \emph{c2d} \citep{Eva09} and {\it Spitzer} Gould Belt\footnote{http://www.cfa.harvard.edu/gouldbelt} Legacy surveys \citep[e.g.,][]{Spe11,Pet11,Hat12,Dun13}.

In this section we provide a similar tool to identify class I/II YSO candidates on the basis of PACS/SPIRE colors. 
Such a tool is extremely useful to identify probable YSOs when no complementary optical/near-IR observations are available, 
and can also be used to estimate the level of stellar contamination in {\it Herschel} extragalactic surveys. However, we stress that this is a very preliminary attempt 
because of the small amount of PACS/SPIRE fluxes for confirmed YSO and galaxy samples published so far. 

We first collected from the literature PACS/SPIRE fluxes for confirmed YSOs. Beside the 29 objects in Cha~II, our final sample includes 49 objects in Cha~I \citep{Win12} and 
28 objects in Serpens, for a total of 106 YSOs. Like Chamaeleon, Serpens is one of the clouds observed within the frame of the HGBS; 
the {\it Herschel} fluxes for the YSOs in this cloud were retrieved from the catalog obtained by \citet{Bre13} and a paper focused on this population is in preparation (Spezzi et al., 2013).
We also used, for comparison purposes, the PACS fluxes of $\sim$50 young very low mass stars and brown dwarfs (BDs) published by \citet{Har12b}, 
and the predicted colors and flux densities of protostars in PACS and SPIRE filters by \citet{Ali10}; these colors are based on a grid of  20160 model SEDs of low-mass protostars obtained by considering emission from 
four main components (a central object, a flared disk, a rotating collapsing envelope and a bipolar cavity) and varying their configuration parameters.

We performed the same literature search for PACS/SPIRE fluxes of extragalactic sources and collected a sample of 96 galaxies:  35 galaxies in the Virgo cluster \citep{Cor12} and 
61 nearby galaxies from the KINGFISH survey \citep{Dal12}. The sample in the Virgo cluster consists of late type star-forming galaxies, from weakly barred spirals (Sab) to irregular galaxies with no bulge component (Sm), 
some of them highly disturbed by the dense cluster environment.
The KINGFISH sample includes galaxies spanning wide ranges in luminosity (over a factor of 10$^4$), optical/infrared ratio (over a factor of 10$^3$), metallicity, gas fraction, H$_I$/H$_2$ ratio, star formation rate, morphology and bar strength; all normal galaxy types are represented and there are several galaxies with nuclei that are clearly distinguished by Seyfert or LINER characteristics, but none of them has an active galactic nucleus (AGN). 
We also used, for comparison purposes i) the galaxy locus defined by \citet{Amb10} on the basis of sources detected by the {\it Herschel}-ATLAS ({\it Herschel} Astrophysical Terahertz Large Area Survey) observations, 
consisting of a flux-limited sample of about 2000 galaxies in the GAMA-9 field near the ecliptic plane, covering a wide range in redshift (0$\lesssim z \lesssim$4 with an average value of 2.2 ) 
and with an average dust temperature of $\sim$28~K; ii) the active AGN locus defined by \citet{Hat10} on the basis of the ``{\it Herschel} Multi-tiered Extragalactic Survey`` (HerMES) observations.

Finally, we estimated the expected colors of main sequence stars and BDs, with no infrared excess emission at {\it Herschel} wavelengths, by convolving the stellar photosphere models by \citet{Hau99} and \citet{All00} with the PACS/SPIRE filter response curves. We considered models with effective temperature between 2000 and 10000~K and  $\log g$=5, as appropriate for main sequence stars and BDs. We then investigate the variation of these synthetic colors as a function of reddening assuming the \citet{Wei01} extinction law for R$_V$= 3.1 (i.e., diffuse interstellar-medium).

\begin{figure*}
\centering
\includegraphics[width=19cm]{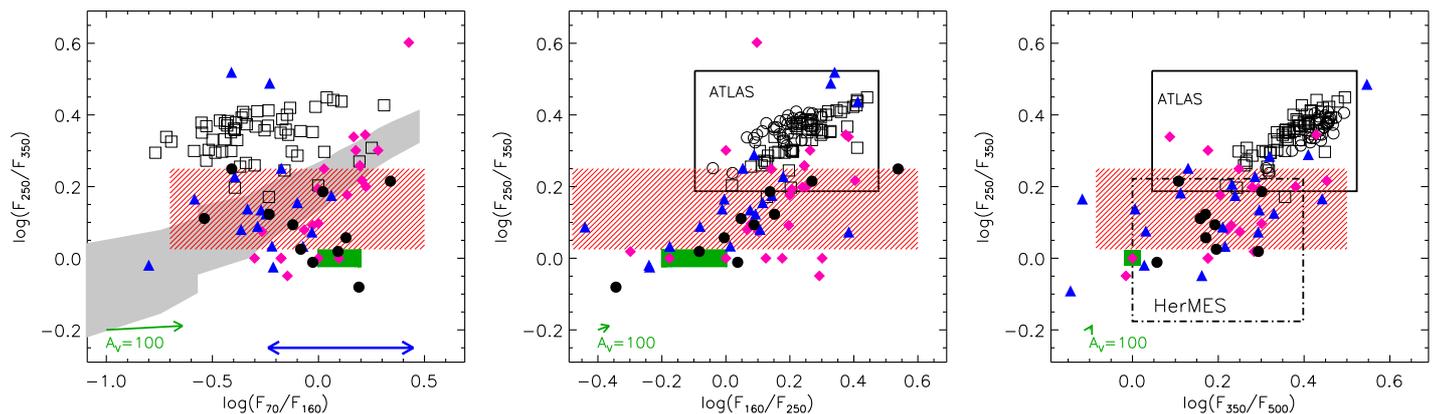}
\caption{PACS/SPIRE color-color diagrams, where colors are defined as the logarithm of the ratio of the flux densities in Jy measured in two different bands. 
The black filled circles are the YSOs in Cha~II, while the magenta diamonds and blue triangles are YSOs in Cha~I \citep{Win12} and Serpens \citep{Bre13}, respectively. 
The red-striped areas mark the YSO loci proposed in this work.
The open squares and circles represent the galaxy samples by \citet{Dal12} and \citet{Cor12}, respectively.
 The green-filled area display the predicted colors for stellar photospheric emission. The arrow indicates the A$_V$=100~mag reddening vector. 
 The solid and dot-dashed lines in the middle and right panels indicate the loci of galaxies in the {\it Herschel}-ATLAS catalog and AGNs in the {\it Herschel}-HerMES survey, respectively. 
 In the left panel, the grey-filled area is the protostar locus predicted by \citet{Ali10}, while the arrow-head line displays the $\log (F_{70} / F_{160})$ color range of the very low-mass stars and brown dwarfs by \citet{Har12b}.}
\label{HGB_diag}
\end{figure*}

We investigate the positions of these YSO, galaxy and reddened photosphere samples on a combination of PACS/SPIRE CC diagrams, 
where colors are defined as the logarithm of the ratio of the flux densities measured in two different bands. 
Figure~\ref{HGB_diag} illustrates three different PACS/SPIRE  CC diagrams. Their inspection shows that 
the YSO population in Cha~II, Cha~I and Serpens share the same color ranges and about 80\% of the YSOs 
in these three clouds are confined in the following locus (indicated by the red-striped areas in the Fig.~\ref{HGB_diag}):

\begin{equation}
-0.7 \lesssim \log (F_{70} / F_{160}) \lesssim 0.5
\end{equation}
\begin{equation}
-0.5\lesssim \log (F_{160} / F_{250}) \lesssim 0.6
\end{equation}
\begin{equation}
0.05 \lesssim \log (F_{250} / F_{350}) \lesssim 0.25
\end{equation}
\begin{equation}
-0.1 \lesssim \log (F_{350} / F_{500}) \lesssim 0.5
\end{equation}

These color ranges are consistent with the PACS/SPIRE colors predicted by \citet{Ali10} for protostars and with the PACS colors 
of the sample of young very low-mass stars and BDs published by \citet{Har12b}. 
PACS/SPIRE colors of reddened photospheres are close to zero regardless of the assumed effective temperature and are unlikely to contaminate the YSO locus defined above, 
which is restricted to $\log (F_{250} / F_{350}) \gtrsim 0.05$, even assuming a very high extinction (i.e., A$_V$=100~mag). 
Galaxies with a broad range of dust temperatures and redshifts \citep{Amb10,Cor12,Dal12} occupy regions in PACS/SPIRE CC 
diagrams which marginally overlap with the YSO locus defined by us, with the expected level of contamination being about 10\%. 
We performed a two-dimensional (2D) Kolmogorov-Smirnov (K-S) test in order to prove that the distribution of YSOs and galaxies in PACS/SPIRE  CC diagrams are significantly different. 
We used the IDL routine {\it ks2d} by P. Yoachim\footnote{http://www.as.utexas.edu/$\sim$yoachim/idl/py\_idl.html}. This routine requires in input  the two 2D arrays to be compared (which are in our case the SPIRE/PACS colors 
of the YSO and galaxy populations), and returns a probability close to 1, or at least not near zero, if the two populations were drawn from same parent distribution, and a probability close to 0 if this is not the case. 
Table~\ref{KS_test} summarizes the results of the K-S test for the CC distribution in the three diagrams presented in Fig.~\ref{HGB_diag}. When comparing  each of the YSO populations in Cha~II, Cha~I and 
Serpens  with the galaxy sample, we find a very low probability, of the order of 1.e-6 or even 1.e-9.
Thus, the K-S test indicates that the distribution of the YSO populations on PACS/SPIRE CC diagrams is significantly different than the distribution of galaxies.

\begin{figure}
\centering
\includegraphics[width=8cm]{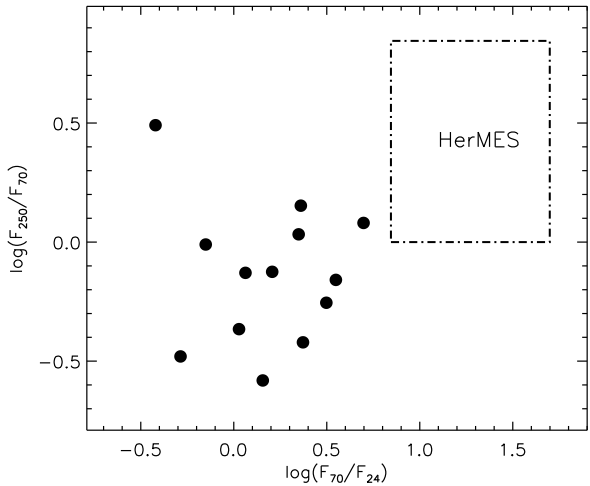}
\caption{$\log (F_{250} / F_{70})$ vs. $\log (F_{70} / F_{24})$ color-color diagram. 
The filled circles are YSOs in Cha~II; their MIPS-24~$\mu$m fluxes were retrieved from the {\it Spitzer}-c2d catalog \citep{Alc08}.
The dot-dashed line indicates the locus of AGNs in the {\it Herschel}-HerMES survey \citep{Hat10}.}
\label{CMD_mips}
\end{figure}

We then conclude that Equations~1-4 provide a reliable YSO selection tool when applied altogether, i.e. for YSO meeting the four color conditions. 
Indeed, Fig.~\ref{HGB_diag} (right panel) also indicate that, when only SPIRE colors are available, YSOs cannot be singled out because 
of the high level of extragalactic contamination, in particular by AGNs, sharing the same SPIRE colors as YSOs.
However, as demonstrated by \citet{Hat10}, AGN samples can be very well separated from the non-AGN, star forming galaxy populations, YSOs, etc., using 
a combination of {\it Spitzer}-MIPS and {\it Herschel}-SPIRE colors, specifically the $\log (F_{250} / F_{70})$ vs. $\log (F_{70} / F_{24})$ CC diagram. 
Indeed, as shown in Figure ~\ref{CMD_mips}, none of the YSOs in Cha~II falls in the AGN area of this diagram.

\begin{table}
\tiny\addtolength{\tabcolsep}{-6pt}
\centering
\caption{Results of the 2D Kolmogorov-Smirnov test comparing the distribution of YSOs and galaxy in PACS/SPIRE CC diagrams (Fig.~\ref{HGB_diag}). }           
\label{KS_test}                           
\begin{tabular}{lccc}      
\hline\hline               
CC diagram   &    Probability$^\dag$  &  Probability$^\dag$  & Probability$^\dag$ \\
                         &     ChaII vs. galaxies &  ChaI vs. galaxies & Serpens vs. galaxies \\
\hline  
$F_{250}/F_{350}$ vs. $F_{70}/F_{160}$       & 6.e-06    & 3.e-09    & 7.e-06 \\
$F_{250}/F_{350}$ vs. $ F_{160}/F_{250}$    &  2.e-07   &  8.e-07   & 6.e-07 \\
$F_{250}/F_{350}$ vs.  $F_{350}/F_{500}$    &  4.e-07    &  7.e-09    & 2.e-08 \\
\hline                                   
\end{tabular}
\\
$^\dag$ The computed probability  is 1 for two populations drawn from same parent distribution, and 0 for two populations with ideally different distributions.\\
\end{table} 

%\begin{table*}
%\tiny
%\centering
%\caption{Results of the 2D Kolmogorov-Smirnov test comparing the distribution of YSOs and galaxy in PACS/SPIRE CC diagrams (Fig.~\ref{HGB_diag}). }           
%\label{KS_test}                           
%\begin{tabular}{lcccccc}      
%\hline\hline               
%CC diagram   &  Probability$^\dag$ &  Probability$^\dag$  & Probability$^\dag$  & Probability$^\dag$  & Probability$^\dag$  & Probability$^\dag$ \\
%                         &  ChaII vs. ChaI  &  ChaII vs. Serpens &  ChaI vs. Serpens &  ChaII vs. galaxies &  ChaI vs. galaxies & Serpens vs. galaxies \\
%\hline  
%$F_{250}/F_{350}$ vs. $F_{70}/F_{160}$       & 0.13 & 0.14 & 0.09 & 6.e-06    & 3.e-09    & 7.e-06 \\
%$F_{250}/F_{350}$ vs. $ F_{160}/F_{250}$    & 0.16 & 0.43 & 0.21 & 2.e-07   &  8.e-07   & 6.e-07 \\
%$F_{250}/F_{350}$ vs.  $F_{350}/F_{500}$    & 0.10 & 0.19 & 0.60 & 4.e-07    &  7.e-09    & 2.e-08 \\
%\hline                                   
%\end{tabular}
%\\
%$^\dag$ The computed probability  is 1 for two populations drawn from same parent distribution, and 0 for two populations with ideally different distributions.\\
%\end{table*} 

\subsection{On the completeness of the YSO and galaxy samples}

The lists of point-like sources observed by the HGBS published so far are limited to previously confirmed 
YSOs \citep[this work;][]{Win12,Rib13}, because the nature of the other detected sources needs further confirmation. 
Thus, there is no systematic study of the photometric completeness of the survey. 
The same issue applies to some of the extragalactic surveys.
Thus, we cannot systematically take into account completeness effects when defining the color ranges of YSOs and galaxies.
However, we give a few warnings about the photometric completeness of the surveys used to collect our 
samples and, hence, the validity range of the YSO locus defined in Sect.~\ref{sec_cc}.

As detailed in Sect.~\ref{sec_cc}, the sample used to define YSO locus was collected from HGBS observations 
of Cha~II, Cha~I and Serpens and the catalog of young very low-mass stars and BDs by \citet{Har12b}.
Although the source extraction was performed in each cloud using different tools, the HGBS observations were conducted in all clouds using the same observing 
strategy and, hence, photometric detection limits are quite similar. 
Our Cha~II catalog has a 5$\sigma$ detection limit of about 150~mJy at 70 and 100~$\mu$m and $\sim$300~mJy at 160, 250, 350 and 500~$\mu$m (see Sect.~\ref{sec_obs} and Fig.~\ref{SNR}-\ref{70micr_comparison}).
\citet{Win12}  adopted the same strategy for source extraction in Cha~I maps \citep[i.e., \emph{getsources};][]{Men12} and 
set a lower limit of 100~mJy in each PACS/SPIRE pass-band for reporting fluxes. 
Using CUTEX \citep{Mol11}, \citet{Bre13} found a point-like source detection limit for Serpens maps of about 300~mJy for PACS and 500~mJy for SPIRE. 
Finally, the catalog of \citet{Har12b} contains an unbiased sample specifically selected to include young objects at or below the sub-stellar limit 
in nearby star forming regions and all of them were detected above a 3$\sigma$ level of a few mJy in PACS pass-bands.
The overall YSO sample span the spectral type range M/K, with only a handful of late G-type objects 
(e.g, see Table~\ref{tab_hgb} of this paper, Table~1 by \citet{Win12} and Table~3 by \citet{Har12b}).

The galaxy sample was collected from several extragalactic surveys conducted with {\it Herschel}.
The sample by \citet{Cor12} in the Virgo cluster is a magnitude-limited group of 35 type Sab to Sm galaxies, 
100\% complete down to 5~Jy with respect to the 250$\mu$m flux. 
\citet{Dal12} report 1$\sigma$ flux limits of 5, 2, 0.7, 0.4 and 0.2 MJy/sr at 70, 160, 250, 350 and 500$\mu$m, respectively, for their sample of 61 nearby galaxies.
The catalog by \cite{Amb10} includes only galaxies detected in at least three of the PACS/SPIRE bands above the 5$\sigma$ 
limit, which varies in the range 35--90 mJy depending on the pass-band. 
Finally, the SPIRE fluxes of AGNs reported by \citet{Hat10} were collected considering all 5$\sigma $ detections at 250 and 350~$\mu $m, corresponding to 12.8 mJy and 12.2 mJy, 
respectively, while no flux cut was applied at 500~$\mu$m.

We conclude that the YSO locus defined in Sect.~\ref{sec_cc} is valid for YSOs of spectral type M/K with 
fluxes above the HGBS 5$\sigma$ limits, which are of the order of 100~mJy and 400~mJy 
for PACS and SPIRE pass-bands, respectively. The locus was defined by taking into account contamination from extragalactic sources 
with fluxes typically higher than 10-100 mJy in the specific pass-bands.

\section{SED fitting and constraints on disk properties \label{sec_fit}}

One of the expected major outcomes from {\it Herschel} observations of class I/II YSOs is a better understanding of their circumstellar disks, 
in particular of the properties and geometry of their cool dust component where most of the mass is located \citep{Har12a}.
SED modeling is the most efficient way to derive these properties when the mid/far IR information comes exclusively from photometry, 
which is the case for the YSO sample in Cha~II. 
However, as pointed out by several authors \citep[e.g.,][]{Rob07,Alc08,Cie11}, disk properties inferred from SED modeling 
are strongly model dependent and degenerate, meaning that 
not only a single set of parameters is consistent with the observed SED. 
As such, they must be treated with extreme caution and used only for statistical purposes 
and to provide typical ranges of disk parameter values.

\citet{Alc08} and \citet{Spe08} presented complete SEDs of the entire pre-main sequence (PMS) populations in Cha~II 
from the optical up to {\it Spitzer} wavelengths, together with reliable estimates of their stellar parameters 
(effective temperature, interstellar extinction, luminosity, radius, mass, age) based on follow-up optical spectroscopy. 
Based on the same dataset, \cite{Alc08} also studied the structure of the circumstellar material around these objects using 
several  sets of SED models for YSOs \citep{Dul01,DAl05,Rob06}. 
Because {\it Spitzer} IRAC and MIPS wavelengths preferentially probe the inner parts of the disk 
and only a few SEDs in their dataset were complete up to millimeter wavelengths, 
\citet{Alc08} could provide constraints only on the inner disk structure 
and found a typical value for the disk inner radius on the order of 0.1~AU.

Merging the \citet{Alc08} and \citet{Spe08} datasets with the new HGBS observations, we have 26 YSOs in Cha~II
with complete SEDs from 0.4-0.5$\mu$m up to 500-1300~$\mu$m, depending on available data, and robust stellar parameter estimates (Fig.~\ref{SEDs}). 
When available, we also included in the SED fluxes at 3.4, 4.6, 12 and 22 $\mu$m from the Wide-field Infrared Survey Explorer \citep[WISE;][]{Wri10}. 
Thus, in the following we attempt to constrain the properties of the cool dust in these 26 disks through SED modeling, 
with the aim of determining which disk parameters are the most likely to be constrained by our new HGBS data, and how. 
Note that, as said above, the HGBS retrieved 29 YSOs in Cha~II.  However, we do not attempt the SED fitting 
for IRAS~12522-7640, STc2d~J130529.0-774140 and IRAS~13036-7644 because of the lack of optical/near-IR data for 
these objects\footnote{IRAS~12522-7640 and IRAS~13036-764 were fist detected by IRAS at mid-to-far IR wavelengths; IRAS~12522-7640 is unclassified, 
while and IRAS~13036-7644  has been reported by \citet{Leh05} to be a transition object between Class 0 and I.  
STc2d~J130529.0-774140 was first detect at mid-IR wavelengths by the {\it Spitzer}-c2d survey \citep{Alc08} \label{note_class0}.}. 

In order to compare our modeling results with the previous modeling by \citet{Alc08}, we explored the same sets of models. 
However, as recently pointed out by \citet{Rob12}, the \citet{Rob06} models are not sufficiently accurate at wavelengths beyond 100$\mu$m and, hence, 
they are not appropriate to reproduce {\it Herschel} data; a new version of these models is currently in preparation \citep{Rob11,Rob12}. 
The model grid by \citet{DAl05} is too coarse and does not allow us to take into 
account the exact stellar parameters of our YSO sample. 
Thus, we used the RADMC-2D code by \citet{Dul04} (Version 3.1, July 2007) to compute SED models for the specific stellar parameters of our YSOs. 
RADMC-2D is an implementation of the one-dimensional semi-analytic models \citep{Dul01} used by \citet{Alc08}.

\subsection{Modeling setup \label{sec_mod_set}}

RADMC-2D is a two-dimensional Monte Carlo code for computing the radiative transfer through an axisymmetric configuration 
of dust \citep{Dul04}\footnote{http://www.mpia-hd.mpg.de/$\sim$dullemon/radtrans/radmc/}. 
The disk is assumed to be passive, i.e. non-accreting, so that the energy balance is entirely determined by the irradiation 
from the central star\footnote{To what extent active accretion changes the stellar flux impinging on the disk surface and, hence, the thermal equilibrium in the disk 
is yet to be determined \citep[e.g.,][]{Woo08}; however, this discussion is beyond the scope of this paper.}. 
Several of the YSOs in Cha~II have a significant mass accretion rate \citep{Spe08,Alc08}; 
however, the presence of mass accretion is expected to affect the SED shape mainly at UV/optical wavelengths, producing an excess emission 
due to the hot gas in the internal accreting envelope, hot spots on the stellar surface heated by the accreting columns, etc. 
\citep{Fei99}, while we use our modeling results only to explore the outer regions of the disk probed by {\it Spitzer/Herschel} wavelengths (Sect.~\ref{sec_mod_res}).

The main input parameters of RADMC-2D are the density structure of the dusty circumstellar material, the dust opacity tables, and the stellar parameters. 
The output is the dust temperature structure, the scattering source functions and a set of SEDs at various inclination angles. 
For each YSO in our sample, we computed a grid of SED models.
Because we do not have mid/far IR spectra of our sources and millimeter fluxes are available only for a few of them, 
we cannot put constraints on the mineralogy and the grain size distribution of the dust and we expect to probe 
dust properties within a few hundred AU from the central star. On the other hand, we have detailed information of the stellar properties and this allows us 
to eliminate some of the degeneracy in the model SEDs. Thus, the following parameters were fixed in all models:

\begin{enumerate}
 
\item For each object, we fixed the stellar mass (M$\star$), radius (R$\star$) and effective temperature 
(T$_{\rm eff}$) to the values spectroscopically determined by \citet{Spe08} 
and reported in Table~\ref{mod_results}. RADMC-2D adopts the Kurucz stellar atmosphere models\footnote{http://www.stsci.edu/hst/observatory/cdbs/k93models.html} 
for the stellar input if the object's T$_{eff} $ is $\ge$3500~K, while for colder objects a blackbody emission is assumed. 
We note that for four of our objects (namely IRAS~12416-7703, ISO-CHAII~28, IRASF~12571-7657 and IRAS~12589-7646) the stellar mass could not be estimated from optical data 
\citep[see Sect.~4.5 and 4.7 by][]{Spe08}, although their spectral type indicates that they are not substellar objects. 
Thus, for these three objects we estimated the stellar mass, together with the disk parameters, from the best fitting SED model, 
considering only models with the same T$_{eff}$ as the object. For IRAS~12500-7658 \citet{Spe08} reported a T$_{\rm eff}$ of 2900~K (see their Sect.~4.7);
however, later follow-up spectroscopy with VLT/FORS2 revealed an earlier spectral type close to K5, 
although the object is strongly veiled (J.M. Alcal\'a, private communication), and this is the value we adopt for our modeling.
 
\item We fixed the slope of the grain size distribution: $dn(a) \propto a^{-3} da$, with minimum grain size a$_{\rm min}$=0.1~$\mu$m and maximum grain size 
a$_{\rm max}$=1~cm. This range is consistent with the values inferred from the analysis of SED slopes at millimeter wavelengths of T~Tauri disks  \citep{Rod06, Lom07,Ric10,Uba12};

\item We assumed the typical chemical composition of dust grains in protoplanetary disks as in \citep{Ric10}, 
i.e. porous composite spherical grains made of astronomical silicates \citep[optical constants from][]{Wei01}, carbonaceous materials \citep{Zub96} 
and water ices \citep{War84}, with fractional abundances from a simplification of the model used in \citet{Pol94} and a volume fraction for vacuum of $\sim$$30\%$; 

\item We assumed a gas-to-dust mass ratio of 100 \citep[][and references therein]{Dut04};
 
\item We fixed the outer disk radius at 200 AU. As shown by \citet{Har12a} and \citet{Cie11}, 
the choice of the outer radius makes essentially no difference to the model SED in the spectral range of our study (typically 0.4-500~$\mu$m).
This is because, for sub-solar mass objects as those in our sample (Table~\ref{mod_results}), most of the IR emission probed by {\it Spitzer}/{\it Herschel} 
data comes from dust at radii inside of $\sim$200~AU;

\item The surface density profile was set to: 
\begin{equation*}
\Sigma(R) = \Sigma_0 \cdot  exp \Bigg[-\Bigg( \frac{R}{R_\mathrm{C}}\Bigg)^{plsig}\Bigg]
\end{equation*}
R$_\mathrm{C}$ is the characteristic radius beyond which the surface density distribution rapidly declines to zero. 
The power law exponent ($plsig$) was set to -1 at radii smaller than R$_\mathrm{C}$ and to -12 at  radii larger R$_\mathrm{C}$.
$\Sigma_0$ is the surface density at R$_\mathrm{C}$ and depends on R$_\mathrm{C}$ and the disk mass (M$_\mathrm{disk}$):  $\Sigma_0= \frac{M_\mathrm{disk}}{2 \pi R_\mathrm{C}^2}$.

\end{enumerate}

Note that these assumptions are consistent with our current understanding of disks around T~Tauri stars \citep[e.g.,][]{Wil11} 
and with the assumptions made by other authors exploring SED modeling including {\it Herschel} photometric data \citep{Cie11,Har12a,Har12b}. 
This makes our results easily comparable with other YSO disk modeling results.

On the other hand, the spectral range covered by our SEDs is expected to probe the inner disk properties through the near/mid IR emission and 
{\it Herschel} observations at longer wavelengths are expected to be more sensitive to the disk mass and degree of flaring \citep[e.g.,][]{Har12a,Har12b}. 
Thus,  for each YSO in our sample, a grid of 2700 SED models was created by varying the following parameters in specific ranges:

\begin{enumerate}

\item We explore seven different values for inner disk radius (R$_\mathrm{in}$):  0.02, 0.05, 0.1, 0.5, 1, 5 and 10 AU. This choice covers the typical range of dust sublimation radii expected 
for T~Tauri stars in our effective temperature range \citep{Wil11} and include also some larger values expected for transition objects with inner holes \citep[e.g.,][]{Muz10,Mer10};

\item We left the pivot point of the density distributions (R$_\mathrm{C}$) vary between 5 and 200 AU.
R$_\mathrm{C}$ is the characteristic radius beyond which the density distribution rapidly declines to zero. Obviously, it must be similar or smaller than the actual disk outer radius (fixed at 200 AU in our case) 
and provides an estimate of the size of the disk region probed by our near-IR to sub-mm photometry, i.e. a lower limit to the actual disk size;

\item  We explore four different values for the disk mass (gas and dust, M$_\mathrm{disk}$): 0.0001, 0.001, 0.01, and 0.1~M$_{\odot}$. 
These values cover the typical range expected for class II YSOs \citep[see Figure~1 by][]{Wil11}. 
Disk masses are well measure at sub-mm/mm wavelengths, where disk are optically thin (i.e., vertical optical depth $\tau <$1) and the dust is virtually all emitting in this range \citep[i.e.,][]{Har12b}.
However, for most of our objects, flux measurement are available up to 500~$\mu$m and disks with the 
typical geometry and range of parameters (R$_\mathrm{in}$, M$_\mathrm{disk}$, a$_{\rm max}$) explored by us, not observed edge-on, 
may still emit a few percent of the total flux at optical depth $\tau >1$ at this wavelength. 
Thus, the M$_\mathrm{disk}$ derived from our SED modeling must be regarded as a lower limit for the actual disk mass;

\item Protoplanetary disk are flared with a vertical scale height (H) that increases with radius. 
We assumed a parametrized power law dependency: $H(R) = H_\mathrm{R_C} \cdot [ \frac{R}{R_\mathrm{C}}]^{1+phi}$. 
H$_\mathrm{R_C}$ is the vertical pressure scale height at R$_\mathrm{C}$ and we set it to 0.15 in unit of radius, a fiducial hydrostatic equilibrium value. 
We left the flaring angle ($phi$) vary between 0.1 and 0.4  in step of 0.1 (i.e. 1.1$\leq 1+phi \leq$1.4), consistently with the analytical/numerical values found by previous studies \citep{Chi97,DAl98,Dul02};
%Note that, depending on the central star, it is not physically meaningful to explore the whole range of flaring angles because we fixed the overall density profile. 
%This is automatically checked by the code and flaring angles not compatible with the assumed density profile are disregarded;

\item We computed each model SED at four inclination angles with respect to the line of sight: 10, 40, 70 and 90 degrees, 
i.e. between the face-on (0 deg) and edge-on (90 deg) configuration.
The inclination angle is poorly determined from SED modeling alone, a well-known result from previous works \citep{DAl99,Chi99,Rob07,Alc08}. 
However, one needs to state the ranges of inclinations that provide a good fit, as there is in some cases a degeneracy between disk size/mass and inclination \citep[][]{Rob07}.

\end{enumerate}

With this configurations of free parameters, we reached a compromise between a sufficiently accurate SED model grid for each YSO and a reasonable computational time. 

 \subsection{Modeling results  \label{sec_mod_res}}

We used an \emph{ad hoc} fitting routine developed under IDL\footnote{Interactive Data Language} 
to select which of the 2700 model SEDs best fits the observed SED for each YSO. 
Prior to the fit, the model grid was scaled to the Cha~II distance (178~pc) and each observed SED was dereddened using the A$_V$ values by \citet{Spe08}, reported in Table~\ref{mod_results}. 

For several of our YSOs only flux upper limits are available at certain wavelengths (see Table~\ref{tab_hgb}) and, hence, our SED fitting has to deal with partly censored data. 
We adopted the likelihood-based approach of \citet{Kel07}, implemented by these authors in the IDL routine {\it linmix\_err.pro}\footnote{http://idlastro.gsfc.nasa.gov/ftp/pro/math/linmix\_err.pro}, 
to determine the best fitting SED model taking simultaneously into account observed flux errors  and upper limits. 
This routine assumes a linear correlation between the dependent and independent variable (i.e., model and observed fluxes, in our case) and performs a Bayesian linear regression to determine the slope, 
the normalization, and the intrinsic scatter of the relationship. The probability model approximates the distribution of the independent variable as a mixture of Gaussians functions. 
Since a direct computation of the posterior distribution is too computationally intensive, random draws from the posterior distribution are obtained using a Markov Chain Monte Carlo method \citep[see Sect.~6.2 by][]{Kel07}.  
The derived likelihood function is modified by including an indicator variable (D), which is equal to 1 for true detections and equal to zero for censored data (i.e., flux upper limits in our case). 
The censored data are taken into account by marginalizing over them when computing the posterior, using the Metropolis-Hastings algorithm \citep{Met53}. 
The intrinsic scatter of the regression ($\varepsilon_0$) is assumed to be normally-distributed and, analogously to the $\chi^2$, provides an estimate of the goodness of fit of the observed distribution to the theoretical one; 
the lower $\varepsilon_0$, the better the quality of the fit is and, hence, the best fitting SED model is determined by minimizing $\varepsilon_0$.

Since SED modeling is known to be highly degenerate, the best-fit model is unlikely to be a unique solution and should be treated with caution.
We therefore adopt the Bayesian method \citep[e.g.,][]{Pin08} to calculate the probability of the disk parameter values sampled by the grid given the available data. 
For each disk parameter, if we assume no prior knowledge on its value, the relative probability of a given model is proportional to $e^{-\varepsilon_0/2}$. 
All probabilities are normalized so the sum of all the probabilities of the models in the grid is equal to 1.
Figures~\ref{fig_prob} and \ref{fig_prob_2} show the probability distribution of each disk parameter for each YSO in the range of values sampled by our grid. 
A rather flat probability distribution indicates that the given parameter its not well constrained (i.e., many/all values sampled by the grid are equally possible).
Clearly, some parameters are better constrained than others, for some objects more than for the others depending on the available SED data. 
In Table~\ref{mod_results} we report the most probable disk parameters (R$_\mathrm{in}$, R$_\mathrm{C}$, M$_\mathrm{disk}$, flaring angle and inclination angle) and the computed $\varepsilon_0$ for the best fitting model;  
when several values of a given disk parameter are equally probable (i.e., partly flat Bayesian probability distribution), we report the compatible range. 

\begin{figure*}
\centering
\includegraphics[height=22.5cm,width=14cm]{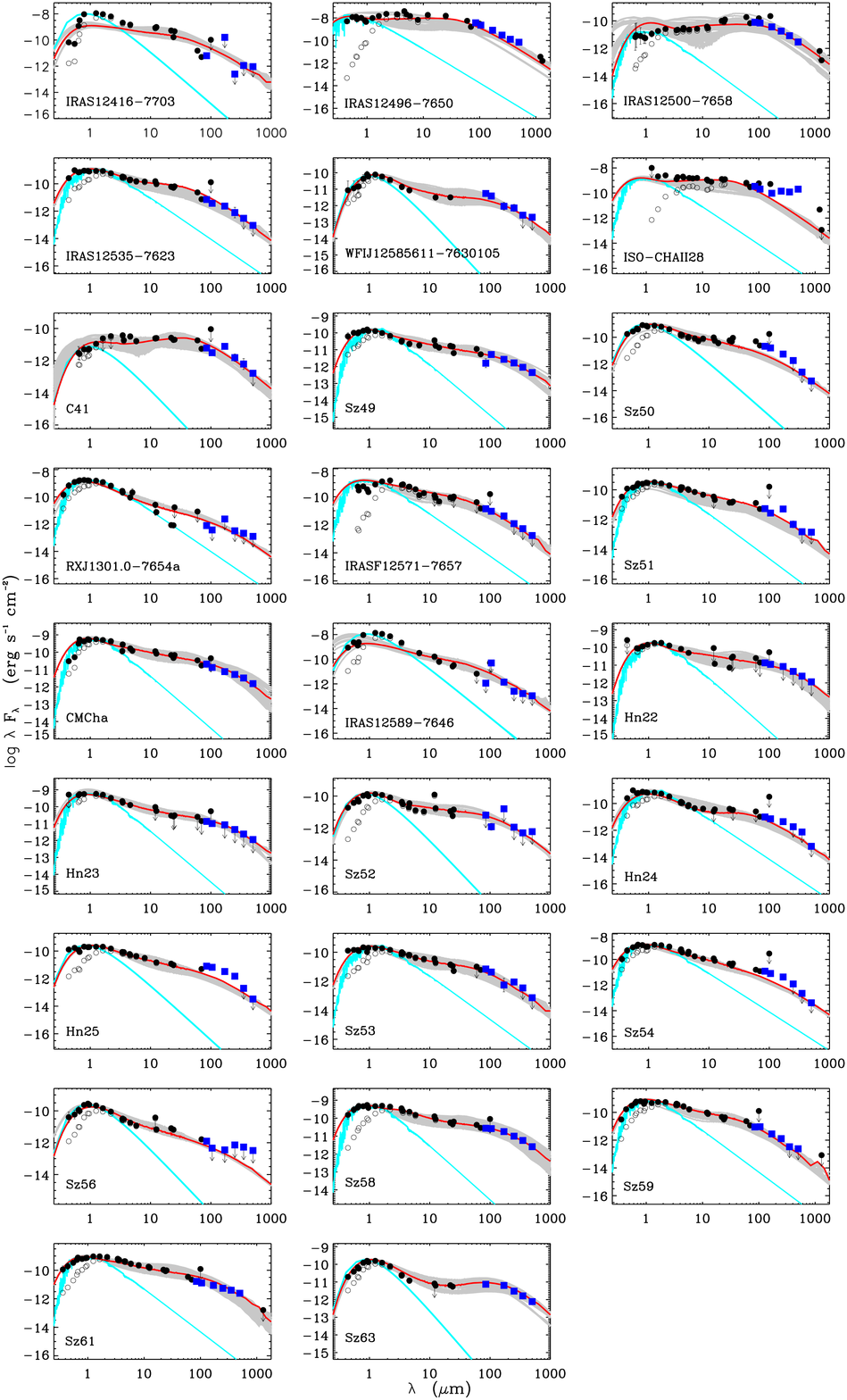}
\caption{Observed (open circles) and dereddened (dots) spectral energy distributions of known YSOs in Cha~II detected by the HGBS. The squares mark PACS and SPIRE fluxes, 
while the dots are optical to infrared flux measurements from previous surveys \citep{Alc08}. 
The Kurucz's model spectrum (for objects with T$_{eff} \ge$3500~K) or the blackbody spectrum (for objects with T$_{eff}<$3500~K) with the same $T_{eff}$ as the object and scaled to its distance and radius \citep{Spe08}, 
is overplotted on each SED, representing the stellar flux. The red thick line is the best fitting RADMC-2D SED model, while the gray lines display the first 100 best-fitting SED models.}
\label{SEDs}
\end{figure*}

\begin{figure*}
\centering
\includegraphics[width=19cm]{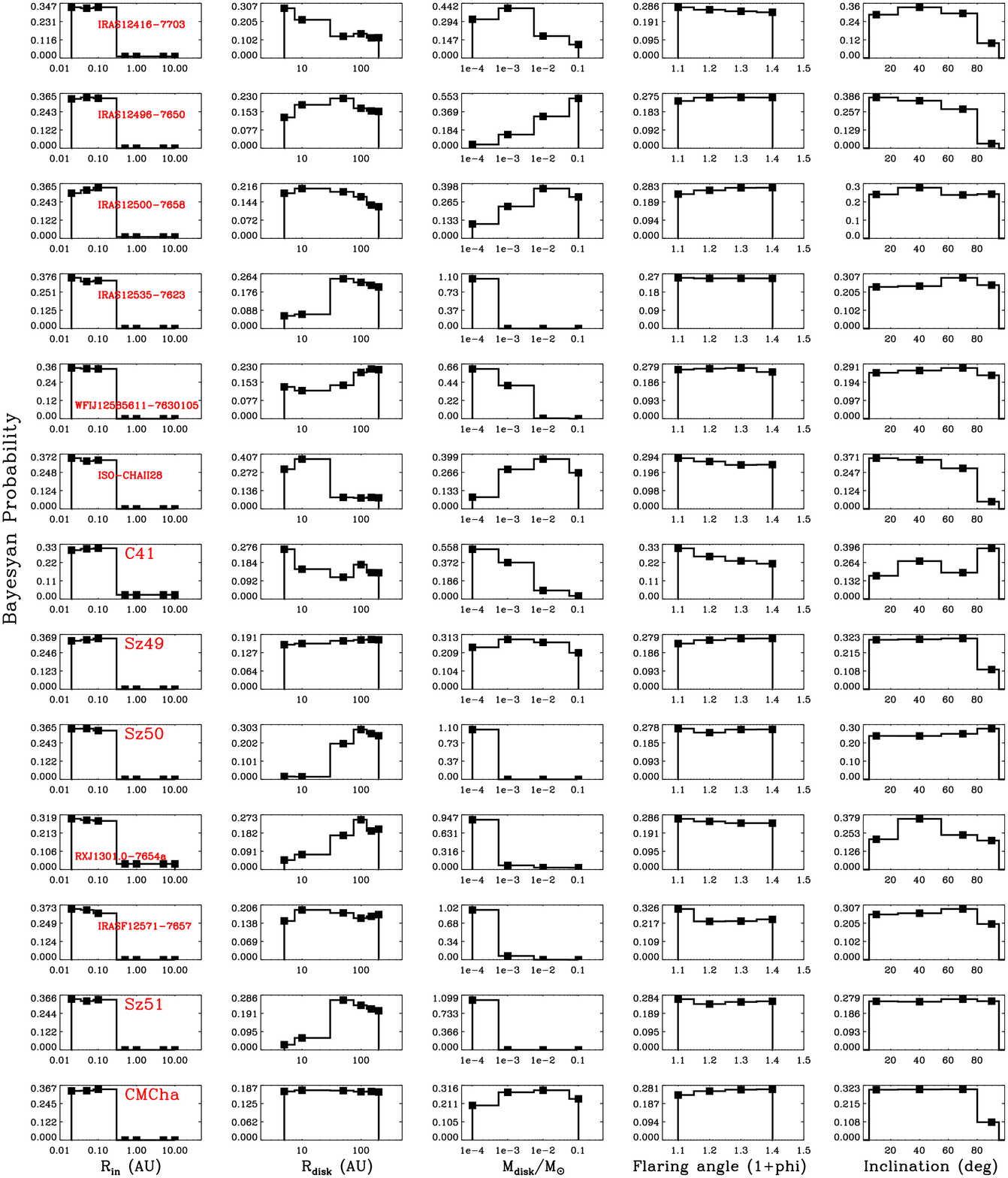}
\caption{Bayesian probability distribution of the five free disk parameters in our SED modeling for the entire SED model grid computed for each YSO. }
\label{fig_prob}
\end{figure*}

\begin{figure*}
\centering
\includegraphics[width=19cm]{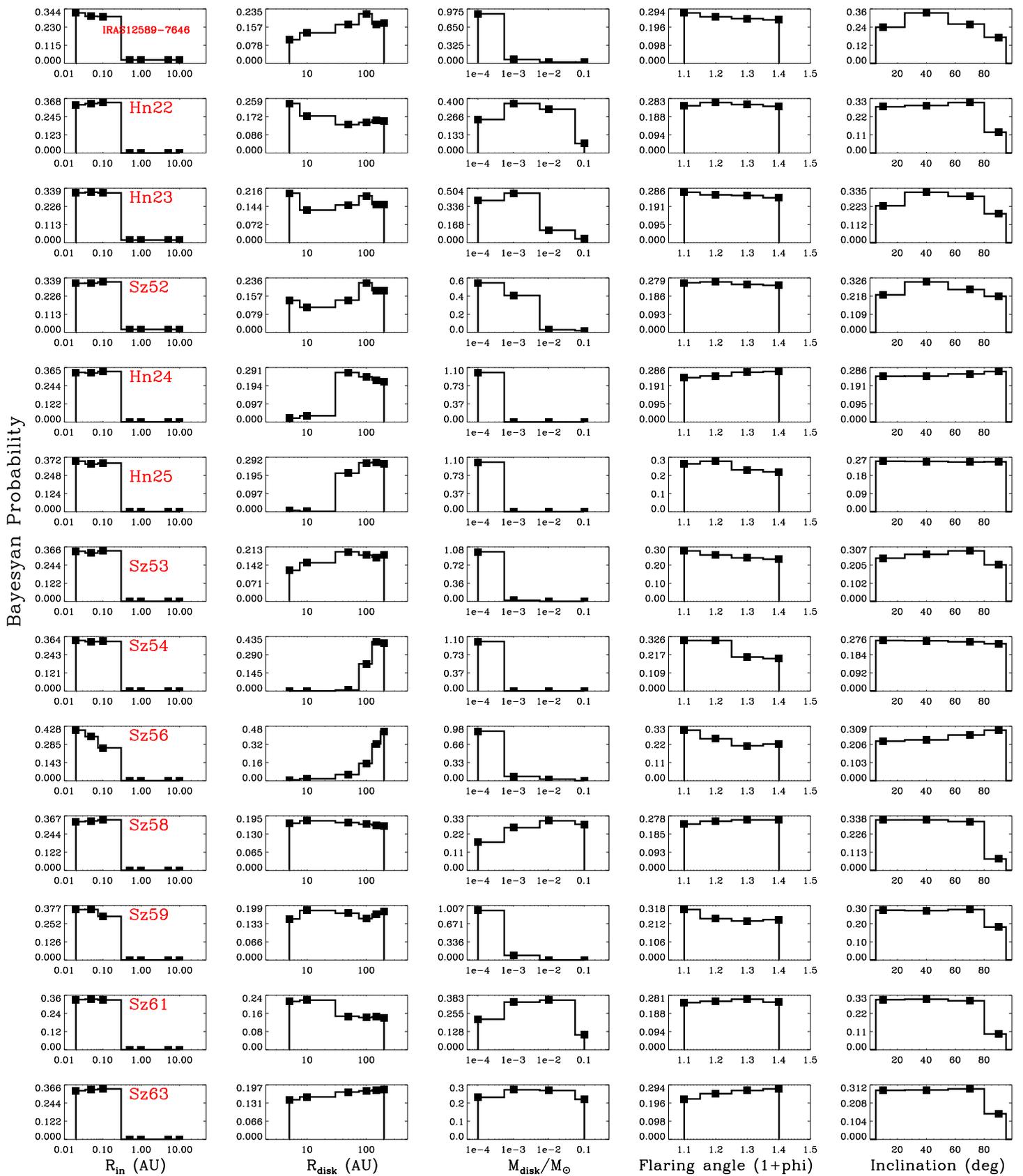}
\caption{Same as Fig.~\ref{fig_prob}. }
\label{fig_prob_2}
\end{figure*}

In Fig.~\ref{SEDs}  we show the SEDs of the 26 YSOs detected in Cha~II by the HGBS together with the best fitting RADMC-2D SED 
model and the first 100 best-fitting SED models (i.e., higher probability models). 
Note that the SED fitting for IRAS~12416-7703, IRAS~12500-7658, and IRAS~12589-7646 is not optimal at optical/near-IR wavelengths ($\lesssim$1~$\mu$m).
IRAS~12500-7658 is a class I source, and IRAS~12416-7703 and IRAS~12589-7646 have 
not been observed through spectroscopy and, hence, their spectral 
types are inferred from broad-band photometry alone \citep{Spe08,Alc08}. 
Thus, the difficulty in modeling the stellar contribution to the SED of these objects probably derive from 
an inaccurate estimate of their stellar parameters (visual extinction and T$_{\rm eff}$), stellar variability and/or 
the presence of an active accreting disk, which affects the total energy balance; 
indeed, this last effect is not taken into account  by RADMC-2D, which assumes a passive disk and irradiation from the central star only.
However, the SED beyond 1~$\mu$m and up to the far-IR is well reproduced by the models and, hence, 
the extracted disk parameters are as reliable as for the other objects (see also discussion in Sect.~\ref{sec_mod_set}). 
For a few objects (ISO-CHAII~28, Sz~50, Hn~25, and Sz~54) the best-fitting SED model appears to underestimate the far-infrared flux ($\lambda \gtrsim 100 \mu$m); 
these objects are located in the region of Cha~II with strongest background emission and, hence, the measured flux densities might be contaminated.
We also note that the SEDs of the Class I object IRAS~12500-7658 is well reproduced at IR wavelength ($\gtrsim$1~$\mu$m) by the disk emission alone, 
without considering the presence of an envelope, normally invoked to model the SEDs of Class I YSOs \citep[e.g.,][]{Lom08}.
 
\begin{table*}
\tiny 
\caption{Stellar properties and most probable disk parameters for the known YSOs in Cha~II detected by the HGBS. For the disk parameters, we indicate between brackets the relative Bayesian probability in percentage.  
When several values of a given disk parameter are equally probable (i.e., partly flat Bayesian probability distribution, Fig.~\ref{fig_prob}-\ref{fig_prob_2}), we report the compatible range.
Disk mass (M$_\mathrm{disk}$) and size (R$_\mathrm{C}$) should be regarded as lower limits (Sect.~\ref{sec_mod_set}). 
In the last column, we report the value of the intrinsic scatter ($\varepsilon_0$) between the observed SED and the best fitting SED model.}           
\label{mod_results}                        
\begin{tabular}{lllllllllll}      
\hline\hline               
Main               &       T$_{\rm eff} ^\dag$   &  A$_V ^\dag$  & R$\star ^\dag$       & M$\star ^\dag$     &  R$_\mathrm{in} ^\clubsuit$ & R$_{C} ^\clubsuit$  & M$_{disk} ^\clubsuit$      &  Flaring angle$^\clubsuit$   & Inclination & $\varepsilon_0$ \\ 
Designation  &         [K]             &  [mag]        & [R$_{\odot}$]        &  [M$_{\odot}$]       &  [AU]         &    [AU]                        &  [M$_\odot$]                                   &   [$1+phi$]       &    angle$^\clubsuit$  [deg]   &   \\
\hline                        

      IRAS~12416-7703 	       &  3487 &  3.36 & 3$^\spadesuit$      &   2$^\spadesuit$  & 0.02-0.1~(32\%)  & $>$5~(27\%)  &  $>$0.001~(40\%)   & 1.1~(26\%)   &  40~(33\%)      	&  1.17  \\
      IRAS~12496-7650	       &  7200 & 10.55 & 2.77		     &   2		 & 0.05-0.1~(33\%)  & $>$50~(20\%)  &  $>$0.1 ~(50\%)   & 1.3-1.4~(25\%)   &  10~(35\%)  	&  0.39  \\
      IRAS~12500-7658$^\star$  &  4350 &  6.85 & 1.00		     &  0.60		 & 0.1~(33\%)  & $>$10~(19\%)  &  $>$0.01~(36\%)   & 1.3-1.4~(26\%)   &  40~(27\%)      	&  0.61  \\
      IRAS~12535-7623 	       &  3850 &  3.36 & 2.71		     &  0.67		 & 0.02~(34\%)  & $>$50~(23\%)  &  $>$0.0001~(99\%)   & 1.1-1.4~(25\%)   &  70~(28\%)   	&  0.18  \\
WFI~J12585611-7630105          &  3025 &  1.66 & 1.13		     &  0.12		 & 0.02-0.1~(33\%)  & $>$150~(20\%)  &  $>$0.0001~(59\%)   & 1.3~(25\%)   &  70~(26\%)  	&  0.07   \\
	 ISO-CHAII~28 	       &  4500 &  38.9 &   2.5$^\spadesuit$  &  1.5$^\spadesuit$ & 0.02~(34\%)  & $>$10~(37\%)  &  $>$0.01~(36\%)   & 1.1~(27\%)   &  10~(34\%)         	&  1.32  \\
		 C~41 	       &  3057 &  2.17 & 0.37		     &  0.10		 & 0.05-0.1~(30\%)  & $>$5~(25\%)  &  $>$0.0001~(50\%)   & 1.1~(30\%)   &  90~(36\%)    	&  0.28 \\
		Sz~49 	       &  3777 &  2.28 & 1.03		     &  0.62		 & 0.05-0.1~(34\%)  & $>$100-200~(17\%)  &  $>$0.001 ~(28\%)   & 1.3-1.4~(25\%)&  10-70~(29\%)  &  0.07  \\
		Sz~50 	       &  3415 &  3.78 & 3.10		     &  0.35		 & 0.02-0.05~(33\%)  & $>$100~(27\%)  &  $>$0.0001~(99\%)   & 1.1-1.4~(25\%)   &  90~(27\%)  	&  0.61  \\
     RX~J1301.0-7654a          &  4350 &  1.93 & 2.67		     &  0.7		 & 0.02~(29\%)  & $>$100~(24\%)  &  $>$0.0001~(86\%)   & 1.1~(26\%)   &  40~(35\%) 		&  0.40   \\
     IRASF~12571-7657 	       &  4730 &  9.03 & 2$^\spadesuit$      &   1$^\spadesuit$  & 0.02~(34\%)  & $>$10~(18\%)  &  $>$0.0001~(92\%)   & 1.1~(30\%)   &  70~(28\%) 		&  0.44   \\
		Sz~51 	       &  3955 &  1.54 & 1.37		     &  0.7		 & 0.02-0.1~(33\%)  & $>$50~(25\%)  &  $>$0.0001~(99\%)   & 1.1~(26\%)   &  10-90~(25\%) 	&  0.35   \\
	       CM~Cha 	       &  4060 &  1.52 & 1.78		     &  0.85		 & 0.1~(34\%)  & $>$10-50~(17\%)  &  $>$0.01~(28\%)   & 1.4~(26\%)   &  10-70~(29\%)  		&  0.18  \\
      IRAS~12589-7646 	       &  3300 &  3.98 & 2.8$^\spadesuit$    &	2$^\spadesuit$   & 0.02~(31\%)  & $>$100~(21\%)  &  $>$0.0001~(88\%)   & 1.1~(27\%)   &  40~(33\%) 		&  0.67   \\
		Hn~22 	       &  3560 &  0.61 & 1.24		     &  0.42		 & 0.05-0.1~(34\%)  & $>$ 5~(23\%)  &  $>$0.001 ~(36\%)   & 1.2~(26\%)   &  70~(30\%)  		&  0.23  \\
		Hn~23          &  4350 &  1.24 & 1.60		     &  1		 & 0.02-0.1~(31\%)  & $>$ 5~(19\%)  &  $>$0.001 ~(45\%)   & 1.1~(26\%)   &  40~(31\%) 		&  0.07   \\
		Sz~52          &  3487 &  4.14 & 1.15		     &  0.35		 & 0.02-0.1~(31\%)  & $>$100~(21\%)  &  $>$0.0001~(54\%)   & 1.2?~(25\%)   &  40~(30\%)  	&  0.25  \\
		Hn~24 	       &  3850 &  2.76 & 2.37		     &  0.65		 & 0.02-0.1~(33\%)  & $>$50~(26\%)  &  $>$0.0001~(99\%)   & 1.4~(26\%)   &  90~(26\%)   	&  0.52 \\
		Hn~25 	       &  3487 &  4.10 & 1.56		     &  0.3		 & 0.02~(34\%)  & $>$150~(26\%)  &  $>$0.0001~(100\%)   & 1.2~(28\%)   &  10-90~(25\%)   	&  0.26 \\
		Sz~53 	       &  3705 &  3.68 & 1.39		     &  0.55		 & 0.02-0.1~(33\%)  & $>$50~(19\%)  &  $>$0.0001~(97\%)   & 1.1~(27\%)   &  70~(28\%)  		&  0.10  \\
		Sz~54 	       &  4350 &  1.57 & 2.42		     &  0.97		 & 0.02~(33\%)  & $>$150~(39\%)  &  $>$0.0001~(99\%)   & 1.1-1.2~(30\%)   &  10-70~(25\%)  	&  0.63  \\
		Sz~56 	       &  3270 &  3.18 & 1.78		     &  0.23		 & 0.02~(39\%)  & $>$200~(44\%)  &  $>$0.0001~(89\%)   & 1.1~(30\%)   &  90~(28\%)   		&  0.37 \\
		Sz~58 	       &  4350 &  3.87 & 1.43		     &  1		 & 0.1~(33\%)  & $>$10~(17\%)  &  $>$0.01~(29\%)   & 1.3-1.4~(25\%)   &  10-40~(31\%)  		&  0.04  \\
		Sz~59 	       &  4060 &  2.67 & 1.96		     &  0.82		 & 0.02-0.05~(34\%)  & $>$10-200~(18\%)  &  $>$0.0001~(91\%)   & 1.1~(29\%)   &  10-70~(27\%)   &  0.28  \\
		Sz~61 	       &  4350 &  3.13 & 1.87		     &  1		 & 0.02-0.1~(33\%)  & $>$10~(21\%)  &  $>$0.01~(34\%)   & 1.3~(26\%)   &  10-70~(30\%) 		&  0.15   \\
		Sz~63 	       &  3415 &  1.61 & 1.38		     &  0.32		 & 0.05-0.1~(33\%)  & $>$200~(17\%)  &  $>$0.001~(27\%)   & 1.4~(27\%)   &  10-70~(28\%) 	&  0.04   \\
\hline
\end{tabular}
\\
$^\dag$ From \citet{Spe08}.\\
$^\clubsuit$ The values sample by the grid are as follows: R$_\mathrm{in}$=[0.02, 0.05, 0.1, 0.5, 1, 5, 10]~AU,  R$_{C}$=[5, 10, 50, 100, 150, 200]~ AU, M$_{disk}$=[0.0001, 0.001,0.01, 0.1]~M$_\odot$, $1+phi$=[1.1, 1.2 ,1.3 ,1.4]  and Inclination=[10, 40, 70, 90]~deg.\\
$^\spadesuit$ Stellar mass and radius not provided by \citet{Spe08}. The values reported here are estimated from the best fitting SED model, 
considering only models with the same T$_{eff}$ as the object.\\
$^\star$ Stellar parameters are estimated assuming spectral type K5 (J.M. Alcal\'a; private communication).\\
\end{table*}

 \subsection{Discussion \label{discuss}}

The estimate of some disk parameters (R$_\mathrm{in}$, R$_\mathrm{C}$ and M$_\mathrm{disk}$) through SED modeling was already attempted 
for the Cha~II sample by \citet{Alc08} on the basis of optical and {\it Spitzer} imaging (see their Table 8 and 9). 
While R$_\mathrm{in}$ is expected to be fairly well constrained by {\it Spitzer} data, the R$_\mathrm{C}$ and M$_\mathrm{disk}$ values derived by us 
are expected to be more accurate than previous estimates, because we used an extended SED dataset which includes {\it Herschel} fluxes up to 500~$\mu$m.
Moreover, thanks to the new HGBS data, we also tried to estimate the degree of flaring ($1+phi$) of these disks, which was never attempted before.

In this section, we discuss in more details the disk parameters derived in Sect.~\ref{sec_mod_res}, comparing them with previous results by \citet{Alc08}  and giving some warning on their use. 
We start by giving a general and very important warning which applies to the use of all the disk parameter values reported in Table~\ref{mod_results}:  
because of the large uncertainty/degeneracy of SED modeling results (outlined in Sect.~\ref{sec_fit}),  
{\it these values must be used as indications of the order of magnitude of each disk parameter for the Cha~II sample and for statistical purposes only, 
while values for individual objects are likely to have large uncertainties}.

\begin{figure*}
\centering
\includegraphics[width=18cm]{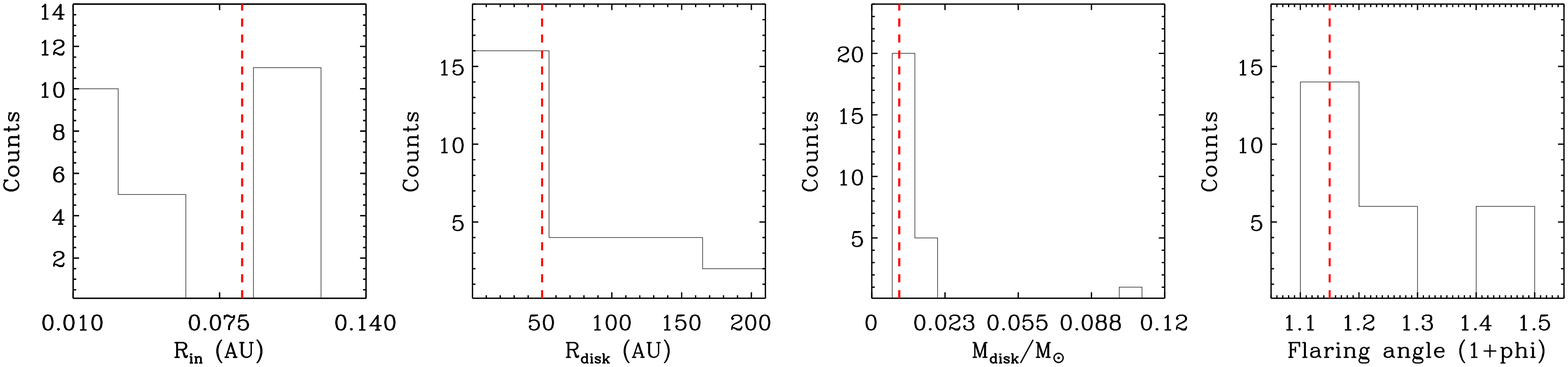}
\caption{Distribution of four free disk parameters in our SED modeling for 26 YSOs in Cha~II detected by the HGBS. We do not show the distribution 
of the disk inclination angles, as it is poorly constrained by the SED modeling (Sect.~\ref{discuss}).
The dashed lines indicates the median value of each distribution.}
\label{histo_diskpar}
\end{figure*}

 {\bf R$_\mathrm{in}$}:  our SED model grid explores seven values for the inner disk radius covering the range of dust sublimation 
radii expected for T~Tauri stars in our T$_{eff}$ range and including larger radii expected for transition objects (Sect.~\ref{sec_mod_set}). 
The choice of this loose grid in R$_\mathrm{in}$ was motivated by the fact that as accurate R$_\mathrm{in}$ 
values as are possible have been already presented by \citet{Alc08} for the Cha~II YSO sample using {\it Spitzer} data and several SED models \citep{Rob06,Dul01,DAl05}.
Thus, we aimed at roughly estimating the range of R$_\mathrm{in}$ values for this sample and confirming the results by \citet{Alc08}  on the basis of 
a SED dataset which is now more complete in the near/mid IR, because it includes WISE and PACS data. 
Considering the many uncertainties in SED modeling, our values are in fairly good agreement with the 
the estimates by \citet{Alc08}, being the relative difference on the order of 30\%.  
We also confirm a typical value for the disk inner radius for the Cha~II sample of the order of 0.1~AU or lower (Fig.~\ref{histo_diskpar}), with a dispersion of 0.04~AU;
considering the stellar radii reported in Table~\ref{mod_results}, this value corresponds to typical inner radii $\sim$7 times larger than the stellar radius. 
We note that, although our grid does not extend beyond 10~AU, none of the 26 objects investigated here has the characteristic SED of transition 
objects with larger inner holes; the Bayesian probability distributions (Fig.~\ref{fig_prob}-\ref{fig_prob_2}) indicate that inner radii  larger than 0.1~AU appear to be inconsistent with the SEDs of all our objects. 
For the majority of them, the R$_\mathrm{in}$ Bayesian probability distribution presents a clear peak, 
indicating that this parameter is fairly well constrained for our sample, as expected because of the good SED coverage at near-to-mid IR wavelengths.
Assuming our R$_\mathrm{in}$ values and using Eq.~11 by \citet{Dul01}, we also estimated the blackbody temperature of the disk inner rim (T$_{rim}$) for our sample; 
depending on the spectral type of the object, T$_{rim}$ spans the range 1000-2500~K, 
with a typical value around 1500~K, consistent with the dust sublimation temperature \citep[typically 900-1600~K for silicates; e.g,][]{Dus96}.

 {\bf R$_\mathrm{C}$}: As clarified in Sect~\ref{sec_mod_set}, R$_\mathrm{C}$ is the characteristic radius beyond which the disk density distribution 
 rapidly declines to zero; it provides an estimate of the size of the disk region probed by our photometry ($\lesssim$500~$\mu$m) and should be regarded as a lower limit to the actual disk size.
The typical value of R$_\mathrm{C}$ for the Cha~II sample is $\sim$50~AU and the dispersion is as large as 60~AU (Fig.~\ref{histo_diskpar}). 
On average, our values are higher than the estimates  provided by \citet{Alc08} and the typical difference is on the order of 50\%. 
This systematic difference probably arises from the better sampling of our SEDs beyond 70~$\mu$m. 
The R$_\mathrm{C}$ Bayesian probability distributions present a clear peak for the majority of the objects (Fig.~\ref{fig_prob}-\ref{fig_prob_2}), supporting the reliability of our estimates.  
We recall the reader that we assumed a power law exponent $plsig$=-1 for the surface density profile at radii smaller than R$_\mathrm{C}$ (Sect.~\ref{sec_mod_set});  a steeper profile  (i.e., greater values of $plsig$) 
would produce larger R$_\mathrm{C}$ \citep[see discussion in][]{Ric13}.

{\bf M$_\mathrm{disk}$}:  As for the disk size, the lack of data at mm-wavelengths for the majority of our objects prevents us from giving a reliable estimate of the total disk mass 
and the values provided in Table~\ref{mod_results} should be regarded as lower limits.  
On the other hand, {\it Herschel} observations at 500~$\mu$m are expected to be sensitive 
to the disk mass \citep[e.g.,][]{Har12a,Har12b} and, indeed, we find a large difference ($\sim$80\%) between our values and the previous estimates by \citet{Alc08}; 
we note, however, that part of this difference is due to different assumptions on the chemical composition of dust grains and, hence, on the dust opacity. 
Our M$_\mathrm{disk}$ Bayesian probability distributions present a clear peak for the majority of the objects (Fig.~\ref{fig_prob}-\ref{fig_prob_2}), supporting the reliability of our estimates.  
The typical value for the Cha~II sample is $\sim$0.0001~M$_{\odot}$ and the dispersion around this value is 0.9~dex (Fig.~\ref{histo_diskpar}). 
Thus, the disk mass distribution in Cha~II appears to be similar to that observed by \citet{And05} in the Taurus-Auriga star formation region 
(typical disk mass 5$\cdot 10^{-3}$~M$_\odot$ and dispersion 0.5 dex) on the basis of a sub-mm survey (350-850$\mu$m) of about 150 YSOs 
in the same stellar mass range, and higher than the median values (3$\cdot 10^{-5}$M$_\odot$) estimated by \citet{Har12b} 
for a sample of $\sim$50 lower-mass stars and BDs on the basis of PACS-only data.
Considering the typical mass of our YSOs (Table~\ref{mod_results}), we find that the lower limits to the disk mass are proportional to the stellar masses with a 0.3\% ratio, 
again very similar to the typical value in Taurus-Auriga \citep[0.5\%;][]{And05} and in the range ($\lesssim$1\%) 
estimated for young Class II stars and BDs across a broad range of stellar masses \citep[0.015-3~M$_\odot$;][]{Kle03,And05,Sch06}. 
This approximately constant disk-to-stellar mass ratio, together with recent observational studies that revealed  stellar-mass dependence for other disk properties,  
such as mass accretion rate \citep[e.g.,][]{Nat04,Muz05}, 
disk lifetime \citep[e.g.,][]{Car06,Pas10} and disk organic chemistry \citep[e.g.,][]{Pas09}, 
indicates that disk properties dependent on stellar properties, 
in contrast with the assumptions of some models of planet formation \citep[see discussion in][]{Szh10,Kor06}.

{\bf Flaring index ($1+phi$)}:  {\it Herschel} observations at far-IR wavelengths (250-500~$\mu$m) are expected to be sensitive to the 
degree of flaring \citep[e.g.,][]{Har12a,Har12b}, and the estimate of this parameter has not been attempted before for the Cha~II sample. 
However, Fig.~\ref{fig_prob}-\ref{fig_prob_2} show that the $1+phi$ Bayesian probability distributions has a tentative trend for some of our YSOs, while is partly flat for others.
Thus, the flaring angle is poorly constrained for our sample; this uncertainty arises from the fact that for several of our YSOs 
only flux upper limits are available in the wavelengths range 250-500~$\mu$m. 
Keeping in mind this limitation, we observe a preferential value around 1.1-1.2, i.e., a low degree of flaring for the great majority of the objects (Fig.~\ref{histo_diskpar}).
In order to understand the meaning of this preferentially low values for the flaring angle, we shall recall that $1+phi$ describes the changing of the pressure 
scale height (H) with the radius (Sect.~\ref{sec_mod_set}); a flatter disk has a smaller H than a flared disk at the same emitting region. Thus, 
lower $1+phi$ values, as the typical ones for our sample, indicate that H increases slower with the radius and, hence, the disk is flatter.
Considering that our objects have spectral types of late K to late M, this result is consistent with the finding by \citet{Szh10} 
in the nearby Chamaeleon~I association based on a sample of 200 G to late M-type stars. 
Although their study is based on {\it Spitzer} IRAC/MIPS data only and, hence, their flaring estimates are even less robust than ours, these authors found indications that disks around lower-mass stars in Chamaeleon~I 
are statistically flatter than those of coeval higher mass stars in the same region. 
Also \citet{Har12b} estimated for their $\sim$50 very low-mass stars and BDs detected by {\it Spitzer} and {\it Herschel}/PACS a typically low flaring angle (1.1-1.2, see their Figure~14). 
This apparent flaring/spectral type anti-correlation has been suggested since several years on the basis of observational studies 
\citep{Apa02,Pas03,Apa05,Szh10}, although disk models predict the opposite \citep{Wal04}, and it further supports the idea 
that disk properties show a dependence on stellar properties.

{\bf Inclination angle}: the disk inclination angle is poorly determined from SED modeling alone \citep{DAl99,Chi99,Rob07,Alc08},
a result confirmed by the rather flat  Bayesian probability distribution observed for the great majority of our YSOs (Fig.~\ref{fig_prob}-\ref{fig_prob_2}).  
Thus, the values reported in Table~\ref{mod_results} must be regarded as 
ranges of inclination providing a good SED fit. As previously mentioned, it is important to state this range, because there is in some cases a degeneracy 
between disk size/mass and disk inclination \citep{Rob07}.

 \subsubsection{Note on Hn~24}

Hn~24 is a class II YSO of spectral type M0 and stellar mass $\sim$0.65~M$_\odot$ (Table~\ref{mod_results}). 
The circumstellar material around this object has been extensively studied by \citet{Mer10}. 
The object was proposed to be a cold disk candidate (i.e., a disk with an inner dust hole larger than the dust sublimation radius) on the basis of its SED, 
which is basically photospheric up to 8-10~$\mu$m and then 
rises at longer wavelengths (see Fig.~\ref{SEDs}). The subsequent spectroscopic follow-up with {\it Spitzer}/IRS revealed the presence of 
10 and 20~$\mu$m silicate features and longer-wavelength crystalline features in the spectrum of Hn~24, though no polycyclic aromatic hydrocarbons (PAHs).
\citet{Mer10} also classified the object as a non-accretor on the basis of the 10\% width of the H$\alpha$ line \citep[i.e.,][]{Nat04} and estimated  a very low disk mass 
($\sim 5 \cdot 10^{-5}~M_\odot$) from SED modeling. All these hints pointed towards an evolved disk around Hn~24. 
However, on the basis of the SED modeling with RADMC \citep{Dul04} and the model grid by \citet{Rob06}, \citet{Mer10} concluded that Hn~24 does not show a significant inner hole.

Our SED modeling with RADMC-2D confirms this result. Models with inner radii larger than 0.5~AU have 
a low Bayesian probability to reproduce the SED of Hn~24 and 
the probability distribution peaks at R$_\mathrm{in} \approx$0.1~AU. \citet{Mer10} estimated an inclination angle around 65~deg for this object, 
consistent with our modeling, which shows that angles larger than 40~deg have a slightly higher probability to reproduce the SED (Fig.~\ref{fig_prob}-\ref{fig_prob_2}). 
Also our disk mass estimate ($> $0.0001~M$_\odot$) is in agreement with the mass inferred by \citet{Mer10}.

\section{Summary and Conclusions \label{concl}}

We used PACS and SPIRE observations of Cha~II performed within the frame of the HGBS key project, complemented by optical/IR imaging and spectroscopy available from the literature, 
to investigate the properties of the Class I to III YSOs in this star forming region.  

We detected 29 out of the 63 known YSOs in Cha~II in at least one of the PACS/SPIRE pass-bands: 3 Class I YSOs (i.e.,100\%), 1 Flat source (i.e., 50\%), 21 Class II objects (i.e., 55\%), 
3 Class III object (i.e, 16\%) and the far-IR source IRAS~12522-7640, not classified because of the lack of near-IR data.

The PACS/SPIRE colors of YSOs  are typically confined to the following ranges, where contamination by other field sources is expected to be low: 
$-0.7 \lesssim \log (F_{70} / F_{160}) \lesssim 0.5$, $-0.5\lesssim \log (F_{160} / F_{250}) \lesssim 0.6$, $0.05 \lesssim \log (F_{250} / F_{350}) \lesssim 0.25$ and 
$-0.1 \lesssim \log (F_{350} / F_{500}) \lesssim 0.5$. When applied altogether, these four color conditions provide a reliable YSO selection tool. 

For 26 YSOs in our sample, we modeled the SED using the RADMC-2D radiative transfer code and analyzed the resulting disk parameter values with a Bayesian method. 
We confirm that the Cha~II YSOs present typical disk inner radii $\lesssim$0.1~AU and, thanks to the new {\it Herschel} data, we put reliable constraints on the 
on the lower limits to the mass and characteristic radius of these disks; we also attempted, for the first time, to estimate their flaring level.
The lower limits to the characteristic radius are typically around 50~AU, although with a large spread, 
and the lower limits to the disk mass are proportional to the stellar masses with a typical 0.3\% ratio. 
The estimated flaring angles, although very uncertain, point towards rather flat disks (1+$phi \lesssim$1.2).

We compared our results with previous estimates in the literature for other samples of low-mass M-type YSOs.
Our results provide further evidence that the disk-to-stellar mass ratio is approximately constant across a broad range of stellar masses and 
that disks around Class II low-mass stars are flatter than disks around coeval higher-mass Class II stars.
This supports the idea that disk properties show a dependence on stellar properties. 
Together with recent studies indicating that the properties of the given star-forming environment (such as metallicity and presence of of strong UV radiation fields) 
may further affect disk evolution \citep{Spe12,DeM11,DeM11b}, our results stress the important unknowns that still subsist in our current theory of disk evolution and planet formation.
This paper paves the way for further studies on these open questions using larger YSOs samples detected by the HGBS in other star forming regions, such as Cha~I and Serpens.

\begin{acknowledgements}

PACS has been developed by a consortium of institutes led by MPE (Germany) and including UVIE (Austria); KU Leuven, CSL, IMEC (Belgium); CEA, LAM (France); MPIA (Germany); INAF-IFSI/OAA/OAP/OAT, LENS, SISSA (Italy); IAC (Spain). This development has been supported by the funding agencies BMVIT (Austria), ESA-PRODEX (Belgium), CEA/CNES (France), DLR (Germany), ASI/INAF (Italy), and CICYT/MCYT (Spain).

SPIRE has been developed by a consortium of institutes led by Cardiff University (UK) and including Univ. Lethbridge (Canada); NAOC (China); CEA, LAM (France); IFSI, Univ. Padua (Italy); IAC (Spain); Stockholm Observatory (Sweden); Imperial College London, RAL, UCL-MSSL, UKATC, Univ. Sussex (UK); and Caltech, JPL, NHSC, Univ. Colorado (USA). This development has been supported by national funding agencies: CSA (Canada); NAOC (China); CEA, CNES, CNRS (France); ASI (Italy); MCINN (Spain); SNSB (Sweden); STFC (UK); and NASA (USA).

We are grateful to Juan M. Alcal\'a for providing the spectral type estimate of IRAS~12500-7658. 
We thank Kees Dullemond for the many explanations on typical parameter ranges for disk around T~Tauri stars and the use of RADMC, 
and Daniel Dale, Luca Cortese and Margherita Bonzini for useful discussions on PACS/SPIRE colors of extragalactic sources.  
We are grateful to an anonymous referee whose suggestions have helped us to improve the presentation of this work. 
NLJC and PR acknowledge support from the Belgian Federal Science Policy Office via the PRODEX Programme of ESA. 
This research has made use of the SIMBAD database, operated at CDS (Strasbourg, France) 
and data products from the Wide-field Infrared Survey Explorer (WISE), which is a joint project of the University of California, 
Los Angeles, and the Jet Propulsion Laboratory/California Institute of Technology, funded by the National Aeronautics and Space Administration

\end{acknowledgements}

\end{document}